\documentclass[twocolumn,showpacs,amsmath,amssymb,prb]{revtex4-1}

\usepackage{graphicx} 
\usepackage{epsfig}
\usepackage{dcolumn} 
\usepackage{color}
\usepackage{changes}
\usepackage{xspace}
\newcommand{\beq}{\begin{equation}}
\newcommand{\eeq}{\end{equation}}
\newcommand{\beqa}{\begin{eqnarray}}
\newcommand{\eeqa}{\end{eqnarray}}

\newcommand{\qvec}{{\bf q}}

\begin{document}

\title{Density inhomogeneities  and Rashba spin-orbit coupling interplay in oxide interfaces}

\author{N. Bovenzi$^1$, S. Caprara$^{2,3}$, M. Grilli$^{2,3,*}$, R. Raimondi$^4$, N. Scopigno$^2$, G. Seibold$^5$}

\affiliation{$^1$Instituut-Lorentz Universiteit Leiden P.O. Box 9506 2300 RA Leiden The Netherlands\\
$^2$ Dipartimento di Fisica Universit\`a di Roma `Sapienza' piazzale Aldo Moro 5 I-00185 Roma Italy \\
$^*$ e-mail: marco.grilli@roma1.infn.it \\
$^3$ Istituto dei Sistemi Complessi CNR and CNISM Unit\`a di Roma Sapienza \\
$^4$ Dipartimento di Matematica e Fisica Universit\`a Roma Tre Via della Vasca Navale 84 00146 Rome, Italy\\
$^5$ Institut f{\"u}r Physik BTU Cottbus-Senftenberg - PBox 101344 D-03013 Cottbus, Germany}

\date{\today}

\begin{abstract}
There is steadily increasing evidence that the two-dimensional electron gas (2DEG) formed at the interface of some
insulating oxides like LaAlO$_3$/SrTiO$_3$ and LaTiO$_3$/SrTiO$_3$ is strongly inhomogeneous. 
The inhomogeneous distribution of electron density is accompanied by an inhomogeneous distribution of the
(self-consistent) electric field confining the electrons at the interface. In turn this inhomogeneous transverse 
electric field induces an inhomogeneous Rashba spin-orbit coupling (RSOC). After an introductory summary on two
mechanisms possibly giving rise to an electronic phase separation accounting for the above inhomogeneity,
we introduce a phenomenological model to describe the density-dependent RSOC and its consequences.
Besides being itself a possible source of inhomogeneity or charge-density waves, the density-dependent RSOC 
gives rise to interesting physical effects like the occurrence of inhomogeneous spin-current distributions and inhomogeneous
quantum-Hall states with  chiral ``edge'' states taking place in the bulk of the 2DEG. The inhomogeneous RSOC can also
be exploited for spintronic devices since it can be used to produce a disorder-robust spin Hall effect.
\end{abstract}

\maketitle

\section{Introduction}
After a two-dimensional electron gas (2DEG) was detected at the interface between two insulating oxides \cite{ohtomo-2004}, 
an increasingly intense theoretical and experimental investigation has been devoted to these systems. The properties of this 
2DEG are intriguing for several reasons. The 2DEG can be made superconducting when its carrier density is tuned by means of 
gate voltage, both in LaAlO$_3$/SrTiO$_3$ (henceforth, LAO/STO) \cite{reyren-2007,caviglia-2008} and LaTiO$_3$/SrTiO$_3$ 
(henceforth, LTO/STO) \cite{biscaras-2010,biscaras-2012} interfaces, thus opening the way to voltage-driven superconducting 
devices. Also, it exhibits magnetic properties\cite{ariando-2010,lilu-2011,bert-2011,dikin-2011,metha-2012,bert-2012}, displays 
a strong and tunable\cite{caviglia-2012,ben-shalom-2010,caprara-2012,hurand-2015} Rashba spin-orbit coupling\cite{rashba-1984}, 
and it is extremely two-dimensional, having a lateral extension $\sim 5$\,nm. Magnetotransport experiments reveal the presence of 
high- and low-mobility carriers in LTO/STO, and superconductivity seems to develop as soon as high-mobility carriers 
appear\cite{biscaras-2012,bell-2009}, when the carrier density is tuned above a threshold value by means of gate voltage, $V_g$. 
When the temperature $T$ is lowered, the electrical resistance is reduced, and signatures of a superconducting fraction are seen 
well above the temperature at which the global zero resistance state is reached (if ever). The superconducting fraction decreases 
with decreasing $V_g$, although a superconducting fraction survives at values of $V_g$ such that the resistance stays finite down 
to the lowest measured temperatures. When $V_g$ is further reduced, the superconducting fraction eventually
disappears, and the 2DEG stays metallic at all temperatures and seems to undergo weak localization at low $T$. At yet smaller 
carrier densities, the system behaves as an insulator. The width of the superconducting transition is anomalously large and it 
cannot be accounted for by reasonable superconducting fluctuations\cite{caprara-2011}. This phenomenology suggests instead that 
an inhomogeneous 2DEG is formed at these oxide interfaces, consisting of superconducting ``puddles'' embedded in a weakly 
localizing metallic background, opening the way to a percolative superconducting transition\cite{bucheli-2013}. Inhomogeneities 
are revealed in various magnetic 
experiments\cite{ariando-2010,lilu-2011,bert-2011,dikin-2011,metha-2012,bert-2012,Prawiroatmodjo-2016}, in tunneling 
spectra\cite{ristic-2011}, and in piezoforce microscopy measurements{\cite{feng-bi-2013}. Specific informations on the doping 
and temperature dependence of the inhomogeneity in these systems have been recently extracted from a theoretical 
analysis\cite{bucheli-2015} of tunnelling experiments\cite{richter-2013}. The inhomogeneous structure of these systems is 
rather complex. On the one hand, large micrometric-scale inhomogeneities have been revealed by the occurrence of striped textures 
in the current distribution\cite{kalisky-2013} and in the surface potential\cite{honig-2013}. On the other hand, experiments 
investigating the quantum critical behavior of the superconductor-(weakly localized) metal transition\cite{biscaras-2013}, 
transport experiments in nanobridges\cite{nanobridges}, and piezo-force experiments\cite{feng-bi-2013} indicate that 
inhomogeneities have a finer structure extending down to nanometric scales. The inhomogeneous character of these oxide 
interfaces (henceforth referred to as LXO/STO interfaces, when referring to both LAO/STO and LTO/STO) has been extensively 
discussed in phenomenological analyses of transport experiments\cite{caprara-2013,caprara-2015}. The very inhomogeneous 
character (especially at small scales) also calls for some intrinsic mechanisms promoting the inhomogeneous distribution of 
electron density. The possibility of an electronic phase separation (EPS) in these materials has indeed been considered and 
two, possibly cooperative, mechanisms have been identified.
In Sect. II, after a short presentation of the electronic structure of LXO/STO interfaces, we will briefly reconsider these 
mechanisms for EPS both for the sake of completeness and to introduce the model that will be the main focus of this
paper: the density-dependent Rashba spin-orbit coupling (RSOC). In this model the RSOC is assumed to depend on the local 
electric field, which in turn is a monotonically increasing function of the local electron density. Therefore, where the 
electron density is larger, also the local confining electric field perpendicular to the interface is larger, thereby inducing 
a stronger RSOC. The subsequent sections will instead be devoted to the analysis of the remarkable consequences of this 
inhomogeneous distribution of electron density and RSOC. 

\section{Two mechanisms for electronic instabilities in LXO/STO}

Photoemission spectroscopy clearly indicates that the valence band of STO and of the LXO overlayer align themselves and the 
excess electrons at the interface are accommodated in the potential well formed by the STO conduction band bending, while 
the conduction band of the overlayer is well above\cite{treske-2015}. This well, which is some tens of
eV deep ($> 0.4$\,eV), gives rise to a quantum confinement of the electrons in the $z$ direction perpendicular to the LXO/STO 
interface  and the interfacial electron gas acquires a strong two-dimensional character. 
Thus the 2DEG resides on the STO side and it occupies the $t_{2g}$ orbitals ($d_{xy},d_{xz},d_{yz}$) of the STO conduction band.
The different orientation and overlap of the orbitals in the ($xy$) and $z$ directions has important consequences in the 
electronic structure of the quantized sub-bands. The $d_{xy}$ orbitals have small overlap along $z$ and give rise to a band with 
small dispersion (heavy mass $m_H\sim 20 m_0$, where $m_0$ is the free electron mass) along this direction. Therefore, when 
quantum confinement is enforced, the sub-band levels are relatively closely spaced. On the other hand, the $d_{xz,yz}$ orbitals 
have a substantial overlap in the $z$ direction and would give rise to dispersed bands (and light masses, $m_L\sim 0.7 m_0$), were 
it not for the confinement. Then the quantized sub-bands are much more widely spaced and the first occupied level is 
$50-100$\,meV above the lowest sub-bands of $d_{xy}$ origin. Both XAS experiments\cite{salluzzo-2009} and first-principle 
calculations \cite{popovic-2008,delugas-2011,zhong-2013} agree on this electronic scheme.

\subsection{Phase separation instability in confined electrons at LXO/STO interfaces}
The thermodynamic stability of the LXO/STO systems was recently investigated\cite{scopigno-2016} in order to identify a 
possible mechanism for EPS. In particular the system was schematized as in Fig.\,\ref{figure1}, where the thin LXO overlayer 
is positively charged because of the countercharges (due to the polarity-catastrophe mechanism\cite{nakagawa-2006,liping-yu-2014} 
and/or to oxygen vacancies) left by the electrons transferred to the STO interface region. These transferred electrons either 
occupy discrete levels in the potential well, which form mobile 2D bands along the $x,y$ directions or are trapped in more 
deeply localized states inside the STO layer.

\begin{figure}[h]
\includegraphics[width=9.5cm]{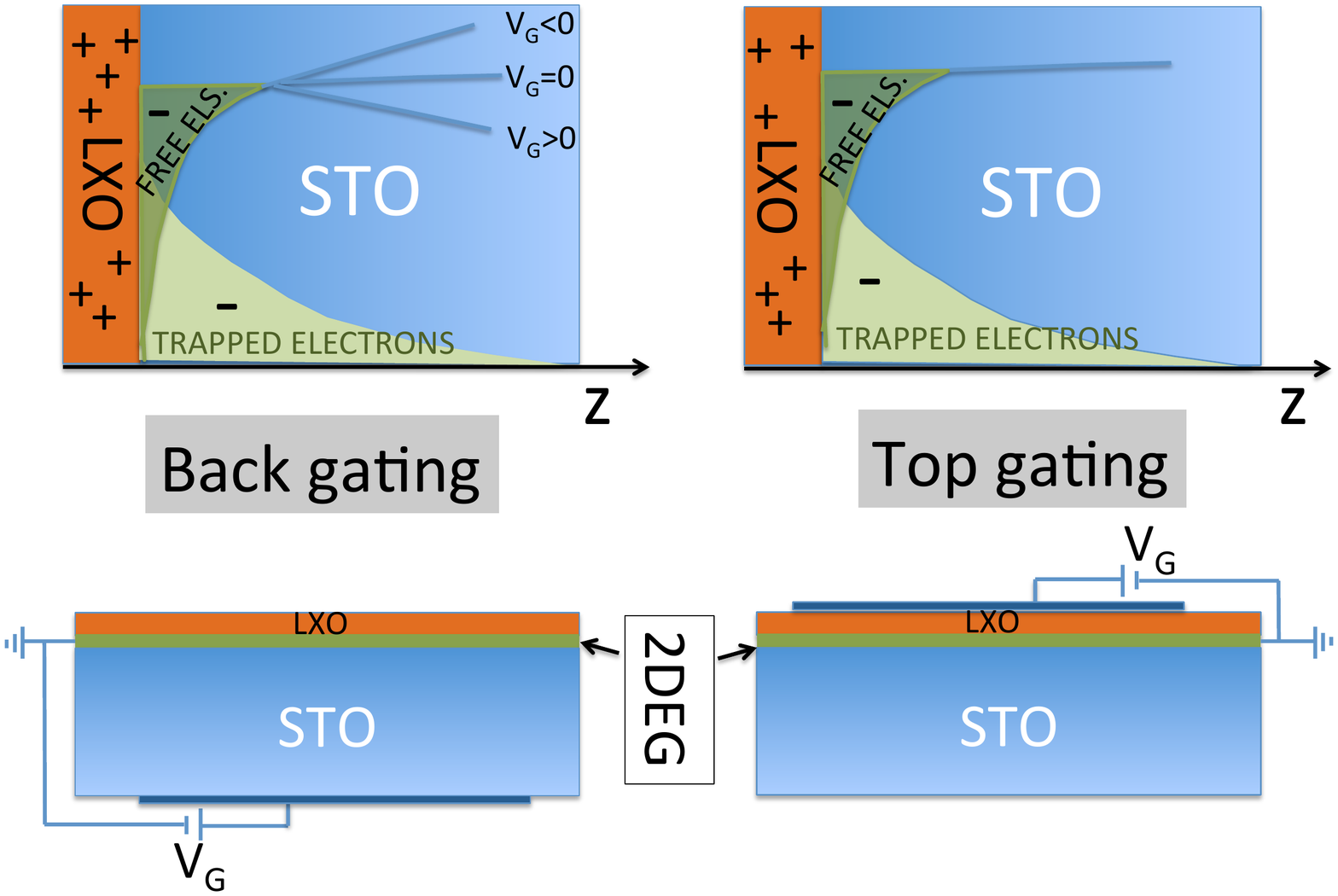}
\caption{Sketch of the interface for back (a) and top (b) gating.
The upper part sketches the confining potentials, while the
bottom part reports the structure of samples and electrodes.
The confining potential depends on both mobile (dark green shade) and trapped
(light green shade) charges, which together compensate the positive 
counter-charges $n$ in the LXO side. Applying a positive (negative)
voltage  electrons are added to (subtracted from) the interface and the potential changes accordingly.}
\label{figure1}
\end{figure}
The electrostatic configuration of the system is also determined by the metallic gates that are under the STO substrate 
(back gating) and/or above the LXO overlayer (top gating), tuning the electron density. The stability of the electronic 
state was investigated by varying the density of the interfacial gas while keeping the overall neutrality. Therefore, a 
corresponding amount of positive countercharges has to be varied (see Fig.\,\ref{figure1}). Because of this tight connection between 
positive and negative charges the (in)stability will be determined by calculating the chemical potential of the whole system 
(i.e., of both the mobile electrons and of the other charges). While we will solve the quantum problem of the mobile
electrons in the self-consistent confining well, the countercharges, the fraction of electrons trapped in impurity states of 
the bulk (see below), and the boundary conditions fixing the gating potential will determine the classical electrostatic energy 
of the system. All these contributions (see Ref.\,\onlinecite{scopigno-2016}) yield the total energy $E$ and, in turn, the chemical 
potential $\mu=E(N+1)-E(N)\approx \partial_N E$ (here $N$ represents the number of electrons, which is always kept equal to the 
number of countercharges). The mobile electron density along $z$ and the spectrum of the discrete levels was determined by solving 
the Schr\"odinger equation along the $z$ direction, while the corresponding electrostatic potential was found by solving the
Poisson equation. By iteratively solving these two equations the self-consistent potential well and the electronic states were 
determined, providing the total energy of the system (also including all electrostatic contributions arising from 
positive countercharges, gate electric fields, trapped localized electrons). From this the chemical potential evolution with 
the electron density $n$ was found, as reported in Fig.\,\ref{figure2}.

\begin{figure}[h]
\includegraphics[width=9cm]{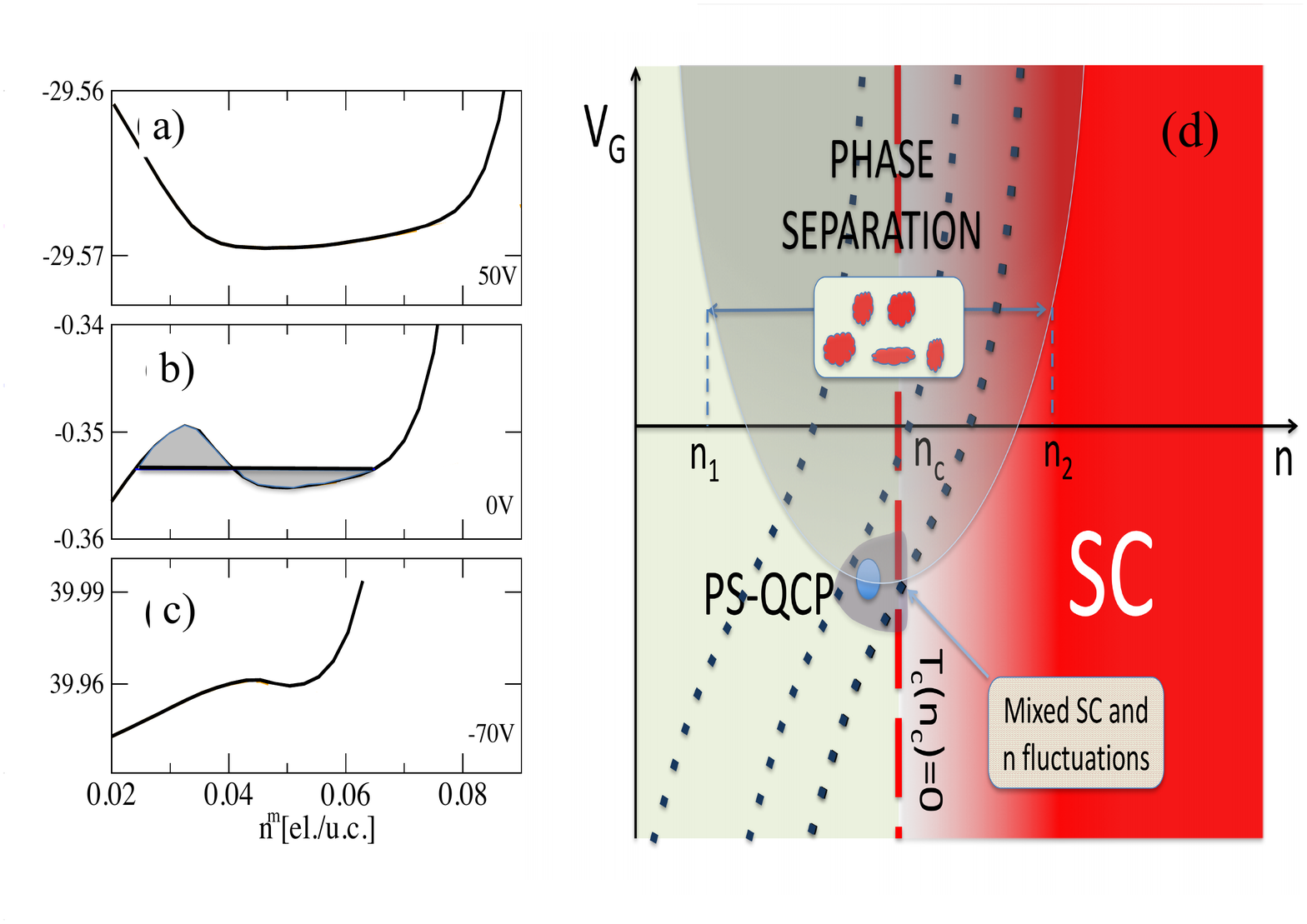}
\caption{(a-c)  Chemical potential as a function of the mobile electron density at fixed values of the back-gating potential 
$V_g$ (the electrons due to gating are thus also fixed) in the presence of a short-range background contribution to the 
chemical potential accounting for the short-range rigidity  of the countercharges  (see Ref.\onlinecite{scopigno-2016})
In (b) an example of Maxwell construction is given, with the gray shaded areas being equal. (d) Sketch of the phase diagram with 
the phase separation region (gray) mixing the superconducting (red) and normal metallic (light green) phases. 
The thick red dashed line marks the critical filling at which SC sets in, while the dotted lines show how the total density 
varies in a back-gating configuration. The darker shaded area marks the 
The densities $n_1$ and $n_2$ delimit the miscibility gap.}
\label{figure2}
\end{figure}
It is clear that at some densities the chemical potential decreases upon increasing $n$, thereby signaling a negative 
compressibility that marks the EPS. The boundaries of the coexistence region are then determined by a standard Maxwell 
construction in full analogy with the liquid-gas transition. As a consequence, a density-vs-gate potential region is determined, 
where regions at different electron density coexist. Of course, the above treatment says nothing about the size of the minority 
droplets embedded in the majority phase: this is determined by specific, model-dependent ingredients like the interface 
energy of the droplets, the mobility of the countercharges that are needed to keep charge neutrality, and so on. 
Simple estimates show that the very large value of the dielectric constant of STO weakens the Coulomb repulsion and allows 
this frustrated EPS mechanism to produce rather large ($\sim 50$ nm) inhomogeneities. On the other hand, it is also possible that 
the positive countercharges (like the oxygen vacancies) diffuse and follow the segregating electrons keeping charge neutrality. 
Of course, also in this case, EPS stops when the segregating electrons become too dense for the countercharges to follow, but 
finite inhomogeneities of substantial size can still be formed. It is important to notice that the calculations find perfectly 
realistic density ranges in which the EPS occurs, with the high-density phases always reaching
local electron densities sufficient to fill the higher orbitals $d_{xz,yz}$. Since these are associated to the high-mobility 
carriers responsible for superconductivity, it is quite natural to assume that the EPS creates puddles at higher-density 
where superconductivity takes place at low-enough temperature. These puddles are then the basic ``bricks'' giving rise to 
the inhomogeneous superconducting state discussed in Sect. I.

\subsection{Phase separation instability in confined electrons with Rashba spin-orbit coupling}
\label{rsoc-eps}

Before the quite effective mechanism for EPS presented in the previous subsection was identified, another mechanism
was found and discussed based on the dependence of the RSOC self-consistent local electric field and, consequently, on the 
local density. Simple inspection of the electrostatic potential well confining the electrons (as obtained from the 
self-consistent Schr\"odinger-Poisson approach) shows that where the electron density is higher, the confining electric field 
is correspondingly higher (on the average in the well). Therefore the RSOC is also larger. Since RSOC brings along a lowering 
of the planar electronic spectrum (for free electrons, if the minimum of the parabolic dispersion is set to zero for $\alpha=0$, 
it becomes $-\varepsilon_0=-\alpha^2m/2$ for finite RSOC), the electron energy tends to be lower in the high-density 
regions. (see Fig.\,\ref{figure3}).

\begin{figure}[h]
\includegraphics[width=8cm]{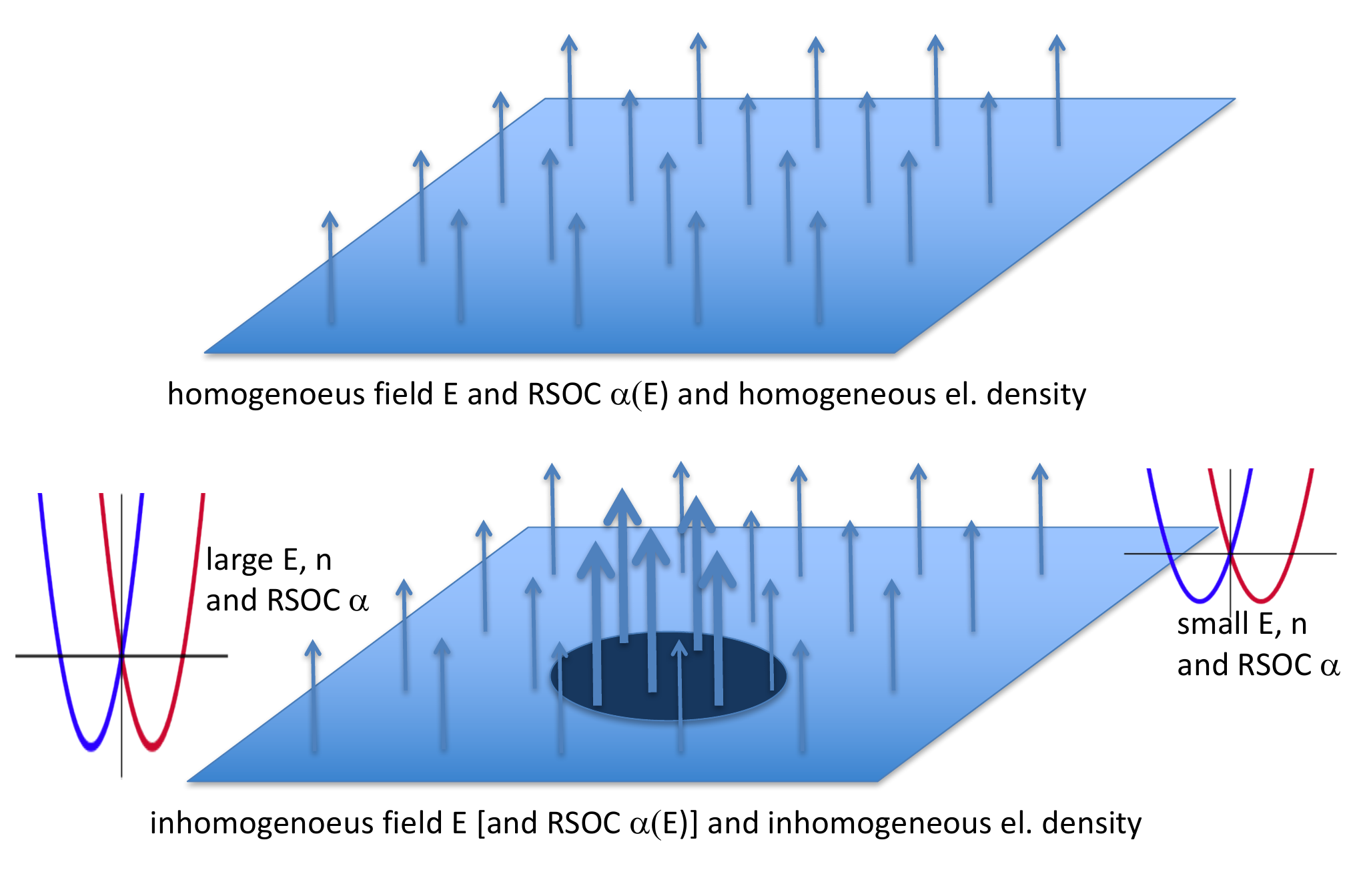}
\caption{Schematic view of the 2DEG in the homogeneous case (above) where both electron density and transverse electric field
(the arrows) are uniformly distributed. (Bottom) Inhomogeneous case where the electric field is larger where the electron density is higher
(darker region). The schematic view of the electron spectrum is reported showing that the bottom of the band is lower when the
RSOC is larger.}
\label{figure3}
\end{figure}
Of course, whether or not this is enough to induce an electronic instability, is a matter of numbers and the detailed 
analysis of this mechanism\cite{caprara-2012,bucheli-2014} established that the DOS of electrons in the dispersive $d_{xy}$ bands 
is not large enough to cause an EPS, while the substantially larger DOS of the higher $d_{xz,yz}$ sub-bands might indeed induce 
this instability for values of the order (or just $30-40$ \% larger) of those experimentally found in LXO/STO 
interfaces\cite{caviglia-2012,ben-shalom-2010,hurand-2015}. After the discovery of the more effective mechanism based on 
the electrostatic confinement (see the previous subsection), the safer attitude is perhaps to consider this mechanism as 
cooperative to strengthen the instability tendency in the 2DEG at the LXO/STO interfaces. Most interestingly, however, is that 
any inhomogeneous distribution of electron density (whatever the formation mechanisms might be) entails an inhomogeneous 
distribution of RSOC, with important consequences both from the fundamental and applicative points of view. The focus of this paper 
is precisely to present some of these important physical and applicative consequences of a density-related
inhomogeneous RSOC.

\section{Rashba model with density dependent coupling}
The basic features resulting from a density dependent RSOC can be elucidated from a single-band Rashba model on a lattice 
described by the hamiltonian

\begin{eqnarray}                                         
H&=&\sum_{ij\sigma}t_{ij}c^\dagger_{i\sigma}c_{j\sigma} + \sum_{ij\sigma\sigma'}
\left\lbrack g^x_{ij} \tau^x_{\sigma\sigma'} + g^y_{ij} \tau^y_{\sigma\sigma'}
\right\rbrack c^\dagger_{i\sigma}c_{j\sigma'}\nonumber \\ &+& \sum_{i,\sigma} \lambda_i \left\lbrack
c^\dagger_{i\sigma}c_{i\sigma} - n_i \right\rbrack
+ \sum_{i\sigma}V_i c^\dagger_{i\sigma}c_{i\sigma}. \label{eq:ham}
\end{eqnarray}   
Here, the first term describes the kinetic energy of electrons on a 
square lattice (with lattice constant $a$)
where we only take hopping between nearest-neighbors into account 
($t_{ij}\equiv -t$ for $|R_i-R_j|=a$). The second term is the RSOC
with $g_{ij}^\alpha = -g_{ji}^\alpha =
-(g_{ij}^\alpha)^*$. Since the coupling constants will be defined as density
dependent, this results  in a local coupling to charge density
fluctuations with $n_i=\sum_\sigma\langle c^\dagger_{i\sigma}c_{i\sigma}\rangle$ and  the $\lambda_i$ are determined 
self-consistently. The last term describes an external (impurity) potential with local
energies $V_i$ which are drawn from a flat distribution with $-V_0 \le V_i \le V_0$.

In our investigations the RSOC is also restricted to nearest-neighbor processes.
We write the couplings as
\begin{eqnarray*}
g^x_{ij} &=& -\mathrm i \gamma_{ij}\left\lbrack \delta_{R_j,R_{i+y}} - \delta_{R_j,R_{i-y}}\right\rbrack \\ 
g^y_{ij} &=& -\mathrm i \gamma_{ij}\left\lbrack \delta_{R_j,R_{i+x}} - \delta_{R_j,R_{i-x}}\right\rbrack 
\end{eqnarray*}
so that the property $g_{ij}^\alpha = -g_{ji}^\alpha$ requires
\begin{equation}\label{eq:gamma}
\gamma_{ij}=\gamma_{ji}.
\end{equation}

The coupling can be also written in the form
\[
H^{RSO}=\sum_i \left\lbrack \gamma_{i,i+y} j_{i,i+y}^x - \gamma_{i,i+x} j_{i,i+x}^y
\right\rbrack
\]
where
\[
j_{i,i+\eta}^\alpha = -\mathrm i\sum_{\sigma\sigma'}\left\lbrack c^\dagger_{i\sigma} \tau^\alpha_{\sigma\sigma'} c_{i+\eta,\sigma'} 
- c^\dagger_{i+\eta,\sigma} \tau^\alpha_{\sigma'\sigma} c_{i,\sigma}
\right\rbrack
\]
denotes the $\alpha$-component of the spin-current flowing on the bond
between $R_i$ and $R_{i+\eta}$. 
Following Ref.\,\onlinecite{caprara-2012}, we assume that the coupling constants
depend on a perpendicular electric field $E {\bf e}_z$ which is proportional
to the local charge density. Since in real space the coupling constants
$g_{i,i+\eta}$ are defined on the bonds, we discretize the electric field
at the midpoints
of the bonds and define the dependence on the charge as
\[
E_{i+\eta/2}=e_0 + e_1 (n_i + n_{i+\eta})\,.
\] 

For the dependence of the RSOC on the electric
field we adopt the form given in Ref.\,\onlinecite{caprara-2012},
so that altogether the following coupling is considered
\begin{equation}\label{eq:coup}
\gamma_{i,i+\eta}= \frac{a_0 + a_1 (n_i + n_{i+\eta})}{\lbrack 1+b_0 + b_1(n_i
+n_{i+\eta})\rbrack^3}\,, 
\end{equation}
which fulfills the property Eq.\,(\ref{eq:gamma}).

We show below that for strong RSOC this coupling will induce the
formation of electronic inhomogeneities and thus concomitant
variations in the local chemical potential $\lambda_i$.
The latter can be obtained self-consistently by minimizing the
energy which yields
\begin{equation}\label{eq:sc}
\lambda_i = \frac{\partial g_{i,i+y}}{\partial n_i} \langle j_{i,i+y}^x\rangle
- \frac{\partial g_{i,i+x}}{\partial n_i} \langle j_{i,i+x}^y\rangle \,.
\end{equation}

\subsection{Stability analysis}
If we insert the Lagrange parameter in the hamiltonian Eq.\,(\ref{eq:ham}) the energy functional reads as
\begin{widetext}
\begin{eqnarray}
E &=& E_0
+\sum_{i\sigma}\left(\frac{\partial\gamma_{i,i+y}}{\partial n_i}\langle 
c^\dagger_{i\sigma}c_{i\sigma}\rangle
+ \frac{\partial\gamma_{i,i+y}}{\partial n_{i+y}}\langle c^\dagger_{i+y,\sigma}c_{i+y,\sigma}
\rangle\right)\langle j_{i,i+y}^x\rangle \nonumber \\
&-&\sum_{i\sigma}\left(\frac{\partial\gamma_{i,i+x}}{\partial n_i}\langle 
c^\dagger_{i\sigma}c_{i\sigma}\rangle
+ \frac{\partial\gamma_{i,i+x}}{\partial n_{i+x}}\langle 
c^\dagger_{i+x,\sigma}c_{i+x,\sigma}\rangle\right)
\langle j_{i,i+x}^y\rangle  \label{eq:exp}
\end{eqnarray}
\end{widetext}
with 
\begin{eqnarray*}
E_0&=&\sum_{ij\sigma}t_{ij}\langle c^\dagger_{i\sigma}c_{j\sigma}\rangle\\
&+& \sum_i \left\lbrack \gamma^0_{i,i+y} \langle j_{i,i+y}^x\rangle - \gamma^0_{i,i+x} 
\langle j_{i,i+x}^y\rangle
\right\rbrack
\end{eqnarray*}
and the notation $\gamma^0=\gamma(n_0)$ refers to the coupling at a given density $n_0$.

From Eq.\,(\ref{eq:exp}) it becomes apparent that the density dependent
coupling induces an effective density-current interaction. 
Assume that the problem has been solved for a given homogeneous density $n_0$
(in the following we take densities and currents as site independent).
Then one can obtain the instabilities of the system from the expansion
of the energy in the small fluctuations of the density matrix in momentum space

\begin{eqnarray}
\delta E&=& Tr(H\delta\rho) \label{eq:deltae}\\ 
&+&\frac{\gamma'}{2N}\sum_q
2\cos(\frac{q_y}{2}) \left\lbrack \delta\rho_q\delta j^x_{-q}
+ \delta j^x_q \delta\rho_{-q}\right\rbrack \nonumber \\
&-&\frac{\gamma'}{2N}\sum_q
2\cos(\frac{q_x}{2}) \left\lbrack \delta\rho_q\delta j^y_{-q}
+ \delta j^y_q \delta\rho_{-q}\right\rbrack \nonumber 
\end{eqnarray}
where $\gamma'$ denotes the (site independent) first derivative of the RSOC with respect to 
the density.

The fluctuations are given by
\begin{eqnarray*}
\delta j^x_{q} &=& -2t\sum_{k\sigma\sigma'} \sin(k_y+\frac{q_y}{2})c^\dagger_{k+q,\sigma}
\tau^x_{\sigma\sigma'}c_{k,\sigma'} \\
\delta j^y_{q} &=& -2t\sum_{k\sigma\sigma'} \sin(k_x+\frac{q_x}{2})c^\dagger_{k+q,\sigma}
\tau^y_{\sigma\sigma'}c_{k,\sigma'} \\
\delta \rho_{q} &=& \sum_{k\sigma\sigma'} c^\dagger_{k+q,\sigma}
{\bf 1}_{\sigma\sigma'}c_{k,\sigma'} 
\end{eqnarray*}
and the instabilities can now be determined from a standard RPA analysis.
We introduce response functions
$$
\chi_{q}(\qvec)=-\frac{\mathrm i}{N} \int dt \langle {\cal T}
\delta A_q(t) \delta A_{-q}(0)\rangle
$$
where $\delta A_q$ refer to the fluctuations defined above.

The non-interacting susceptibilities can be obtained from the
eigenstates of the Rashba hamiltonian Eq.\,(\ref{eq:ham}). 
Denoting the response functions in matrix form
\begin{displaymath}
\underline{\underline{\chi^{0}({\bf q})}}=
\left(\begin{array}{ccc}
\chi^{0}_{jx,jx} & \chi^{0}_{jx,jy} & \chi^{0}_{jx,\rho} \\
~&~&~\\
\chi^{0}_{jy,jx} & \chi^{0}_{jy,jy} & \chi^{0}_{jy,\rho} \\
~&~&~\\
\chi^{0}_{\rho,jx} & \chi^{0}_{\rho,jy} & \chi^{0}_{\rho,\rho}
\end{array}\right)
\end{displaymath}
and the interaction, derived from Eq.\,(\ref{eq:deltae}) as
\begin{displaymath}
\underline{\underline{V({\bf q})}}=
\left(\begin{array}{ccc}
0 & 0 & 2\gamma'\cos(\frac{q_y}{2}) \\
~&~&~\\
0 & 0 & -2\gamma'\cos(\frac{q_x}{2}) \\
~&~&~\\
2\gamma'\cos(\frac{q_y}{2}) & -2\gamma'\cos(\frac{q_x}{2}) & 0
\end{array}\right)
\end{displaymath}
the full response is given by
\begin{equation}
\underline{\underline{\chi({\bf q})}}=\left(\underline{\underline{\bf 1}} 
- \underline{\underline{\chi^{0}({\bf q})}}
\underline{\underline{V({\bf q})}}\right)^{-1}
\underline{\underline{\chi^{0}({\bf q})}}
\end{equation}
and the instabilities can be obtained from the zeros of the determinant
\begin{displaymath}
\left|\underline{\underline{\bf 1}} - \underline{\underline{\chi^{0}({\bf q})}}
\underline{\underline{V({\bf q})}}\right| = 0 \,.
\end{displaymath}
\begin{figure}[htb]
\includegraphics[width=8cm,clip=true]{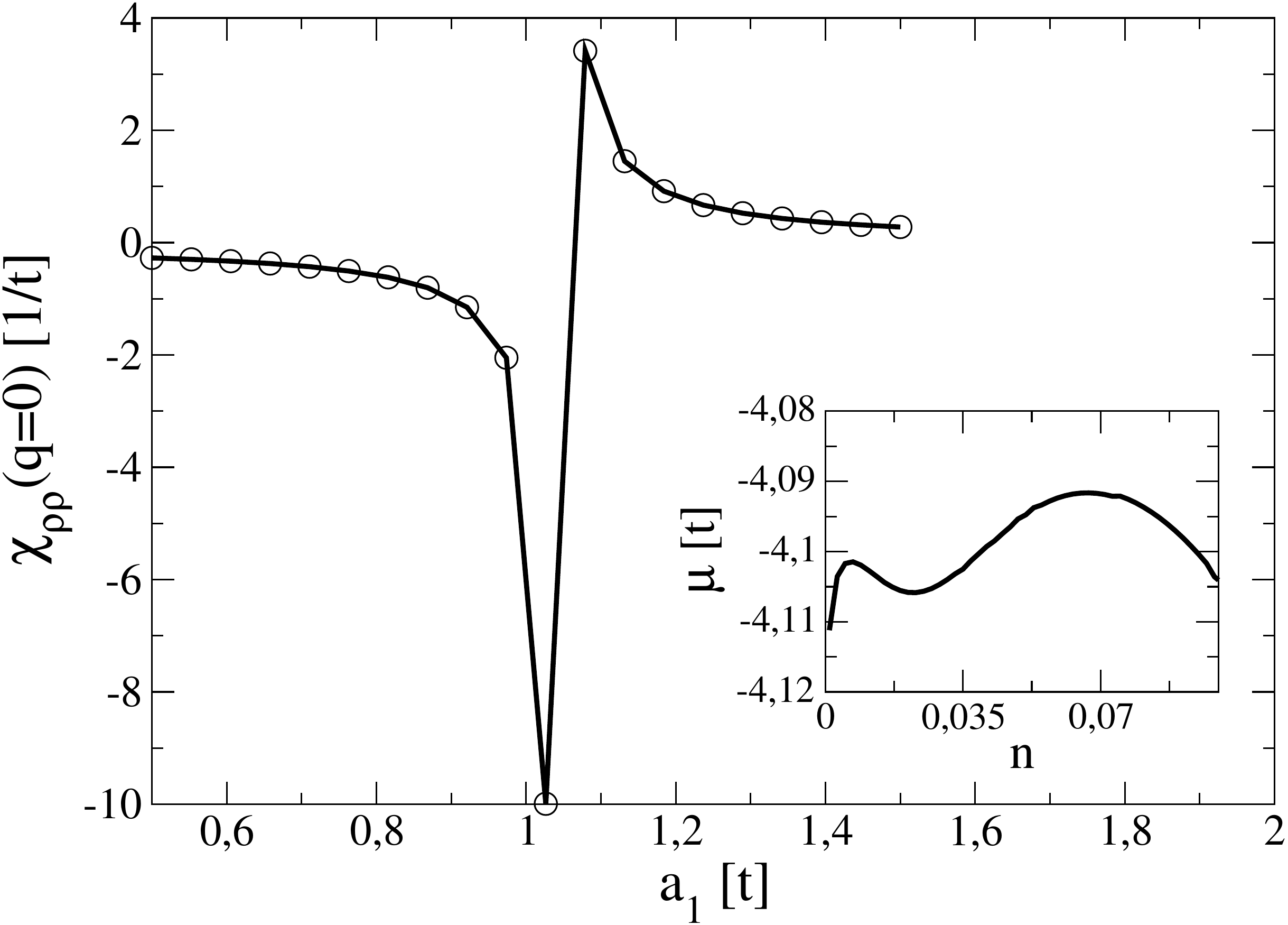}
\caption{Main panel: $\chi_{\rho\rho}(q=0)$ vs. $a_1$ for $a_0=0.3$ and
density $n=0.07$. A $q=0$ instability occurs at $a_1 \approx 1.05$. 
Inset: The $\mu$ vs. $n$ curve for parameters $a_0=0.3$ and $a_1=1.05$
which demonstrates the zero slope at $n=0.07$.}
\label{fig12}
\end{figure}
Here the element $\chi_{\rho\rho}(q=0)$ is proportional to the compressibility, 
i.e. within our sign convention proportional to the inverse of 
$-\partial\mu/\partial n$. A (locally) stable system thus corresponds
to $\chi_{\rho\rho}(q=0)<0$ whereas an unstable system is characterized by
$\chi_{\rho\rho}(q=0)>0$. 

Fig.\,\ref{fig12} demonstrates the consistency of the present approach.
For density $n=0.07$ and fixed $a_0=0.3$ the main panel shows 
$\chi_{\rho\rho}(q=0)$ as a function of  $a_1$. Obviously the system
changes from locally stable to locally unstable at $a_1\approx 1.05$.
This is consistent with the $\mu$ vs. $n$ curve which is shown in the
inset and shows a zero slope for the same parameters (cf. also upper
left panel of Fig.\,\ref{fig1}).

Due to the momentum dependence of the density-current coupling and
the momentum structure of $\chi^0(q)$ a finite $q$ instability can
occur before. This is demonstrated in the left panel of Fig.\,\ref{fig15} 
which shows the momentum dependence of $\chi_{\rho\rho}(q)$ along the
$x$-direction. Clearly, as a function of $a_1$  a instability occurs 
at ${\bf q} \approx (0.3,0)$ before the ${\bf q} =0 $ instability
is reached. Moreover, the corresponding momentum is larger than one would
expect from the nesting momentum of the upper band which is shown
in the right panel of Fig.\,\ref{fig15} which clearly reveals the
importance of the momentum dependent coupling.
A detailed investigation of the phase diagram and the corresponding
structure of instabilities as a function of doping is presented
in Ref.\,\onlinecite{epl15ps}. In this latter work it was also shown that 
the Maxwell construction establishing the whole phase separated region
preempts reaching the finite-$q$ instability

\begin{figure}[htb]
\includegraphics[width=8cm,clip=true]{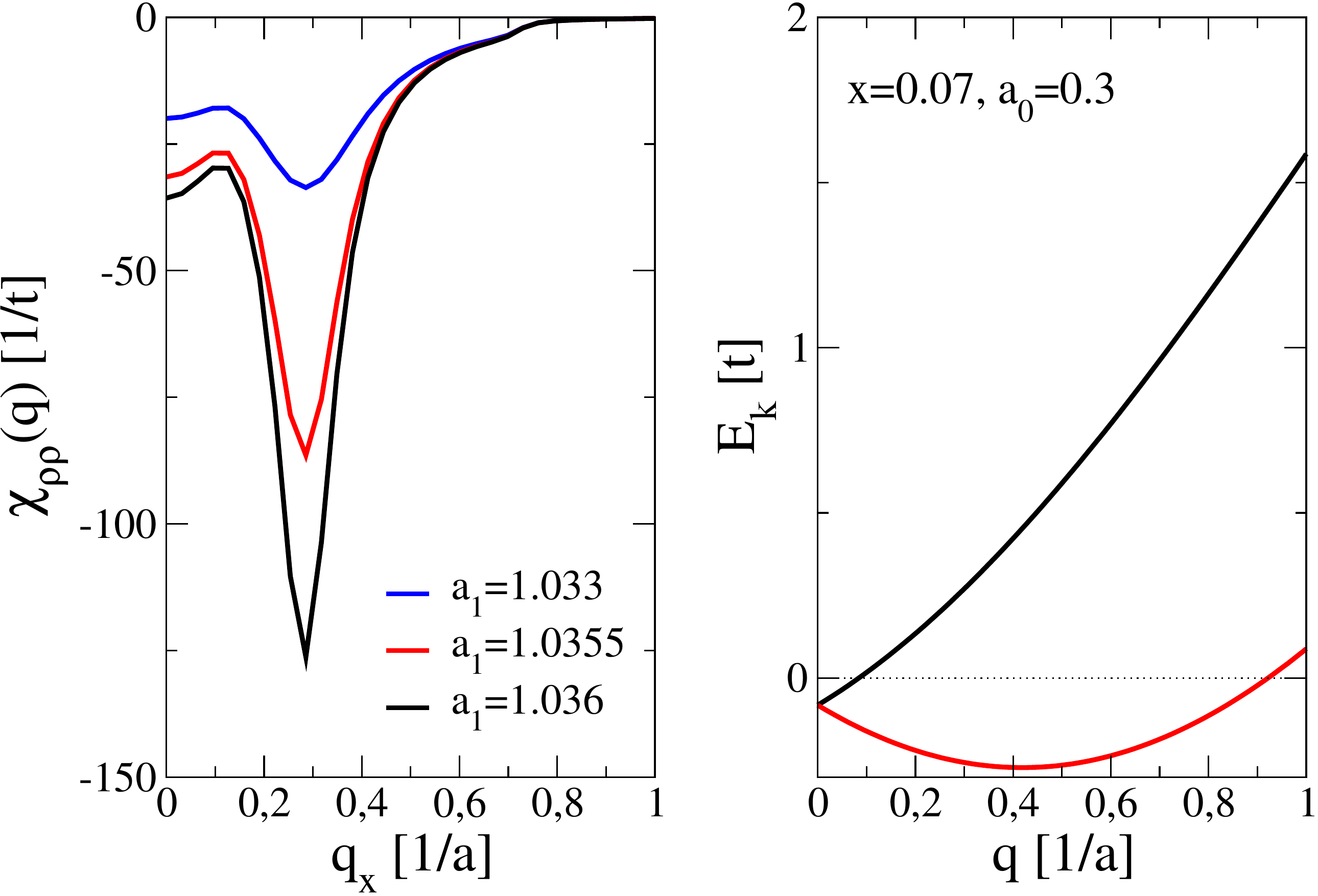}
\caption{Left panel: $\chi_{\rho\rho}(q)$ vs. $(q_x,0)$
for several $a_1$ ( momenta $q_x$ are in units of inverse of the lattice spacing $a$).
Right panel: band structure along $(q_x,0)$.
Parameters:  $a_0=0.3$ and density $n=0.07$. }
\label{fig15}
\end{figure}

\subsection{Spin currents}
Spin currents and associated torques are important quantities in characterizing
the ground state of inhomogeneous Rashba models.
In fact, the electron spin ${\bf S}$ is not a conserved quantity in systems
with spin-orbit coupling. It obeys the Heisenberg equation of motion
\begin{equation}
\frac{d{\bf S}}{dt} = -\mathrm i \lbrack {\bf S}, H\rbrack + \frac{\partial 
{\bf S}}{\partial t}
\end{equation}
which can be interpreted in terms of a continuity equation
\begin{equation}
{\bf G} = \mathrm{div}\, {\bf J} + \frac{\partial {\bf S}}{\partial t}
\end{equation}
where ${\bf G}$ is a 'source' term which in general is finite
due to the non-conservation of spin.
Since we are dealing with the time-independent Schr\"odinger equation
where all expectation values are stationary, the source term is
 'hidden' in the commutator, i.e.
\begin{equation}
\lbrack {\bf S}, H\rbrack = \mathrm i \,\mathrm{div}\, {\bf J} -\mathrm i {\bf G} 
\end{equation}
and ${\bf G}$ contains all contributions which cannot be associated
with a divergence.

In particular one obtains for $J^z$ and $G^{x,y}$
\begin{eqnarray*}
J^z_{i,i+x(y)} &=& -\mathrm i \sum_{\sigma\sigma'}\left\lbrack c^\dagger_{i\sigma} 
\tau^z_{\sigma\sigma'} c_{i+x(y),\sigma'} - c^\dagger_{i+x(y),\sigma} 
\tau^z_{\sigma'\sigma} c_{i,\sigma}\right\rbrack \label{eq:jz}\\
G_i^{x(y)} &=& \mathrm i \gamma_{i,i+x(y)}\times \\
&\times&\sum_{\sigma\sigma'} \left\lbrack c_{n\sigma}^\dagger 
\tau^z_{\sigma\sigma'}c_{i+x(y)\sigma'} - c_{i+x(y)\sigma}^\dagger \tau^z_{\sigma\sigma'}c_{i\sigma'}\right\rbrack
\end{eqnarray*}
corresponding to the relations
\begin{equation}\label{eq:gj}
G_i^{x(y)}=-\gamma_{i,i+x(y)} J^z_{i,i+x(y)} - \gamma_{i-x(y),i} J^z_{i-x(y),i},
\end{equation}
i.e., a torque for the $x(y)$-component of the spin is associated with
a $z$-polarized spin current along the $x(y)$ direction when the RSOC $\gamma \ne 0$.

In a system with homogeneous RSOC one has finite $x(y)$-polarized spin currents
flowing along the $y(x)$-direction. In particular, since the currents are
constant the corresponding torques vanish and from Eq.\,(\ref{eq:gj})
it turns out that $z$-polarized spin currents are absent in the
homogeneous system. In the next subsection we demonstrate that the
situation drastically changes when the RSOC depends on the density
and thus induces an inhomogeneous charge distribution in the ground state.

\begin{figure}[thb]
\includegraphics[width=8cm,clip=true]{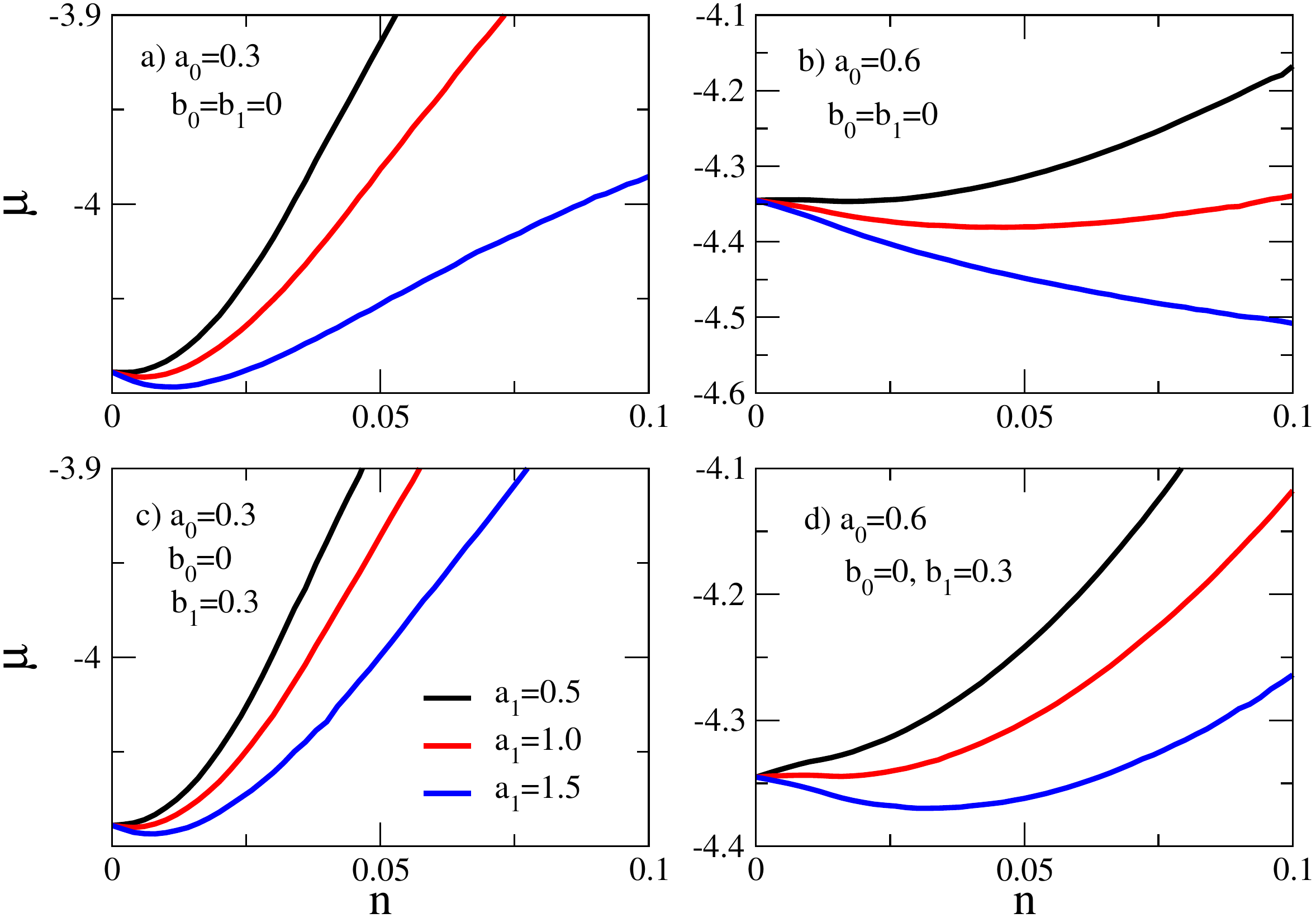}
\caption{Chemical potential {\it vs} charge density for
different parameters of the coupling constant.
Each panel shows the curves for $a_1=0.5$ (black), $a_1=1.0$ (red),
and $a_1=1.5$ (blue), whereas $a_0$ and $b_1$ are different in each panel.}
\label{fig1}
\end{figure}

\section{Results for inhomogeneous charge and spin structures}
In case of a homogeneous system Fig.\,\ref{fig1} displays the 
chemical potential {\it vs} density
for the various parameters entering the coupling constant Eq.\,(\ref{eq:coup}).
For simplicity only the case $b_0=0$ in Eq.\,(\ref{eq:coup}) is considered. 
Clearly a phase separation instability is triggered by increasing the
RSOC to the density via the parameter $a_1$. On the other hand the 
parameter $b_1$ puts an upper limit to this coupling so that the PS
instability is shifted to lower doping upon increasing $b_1$.
\begin{figure}[htb]
\includegraphics[width=8cm,clip=true]{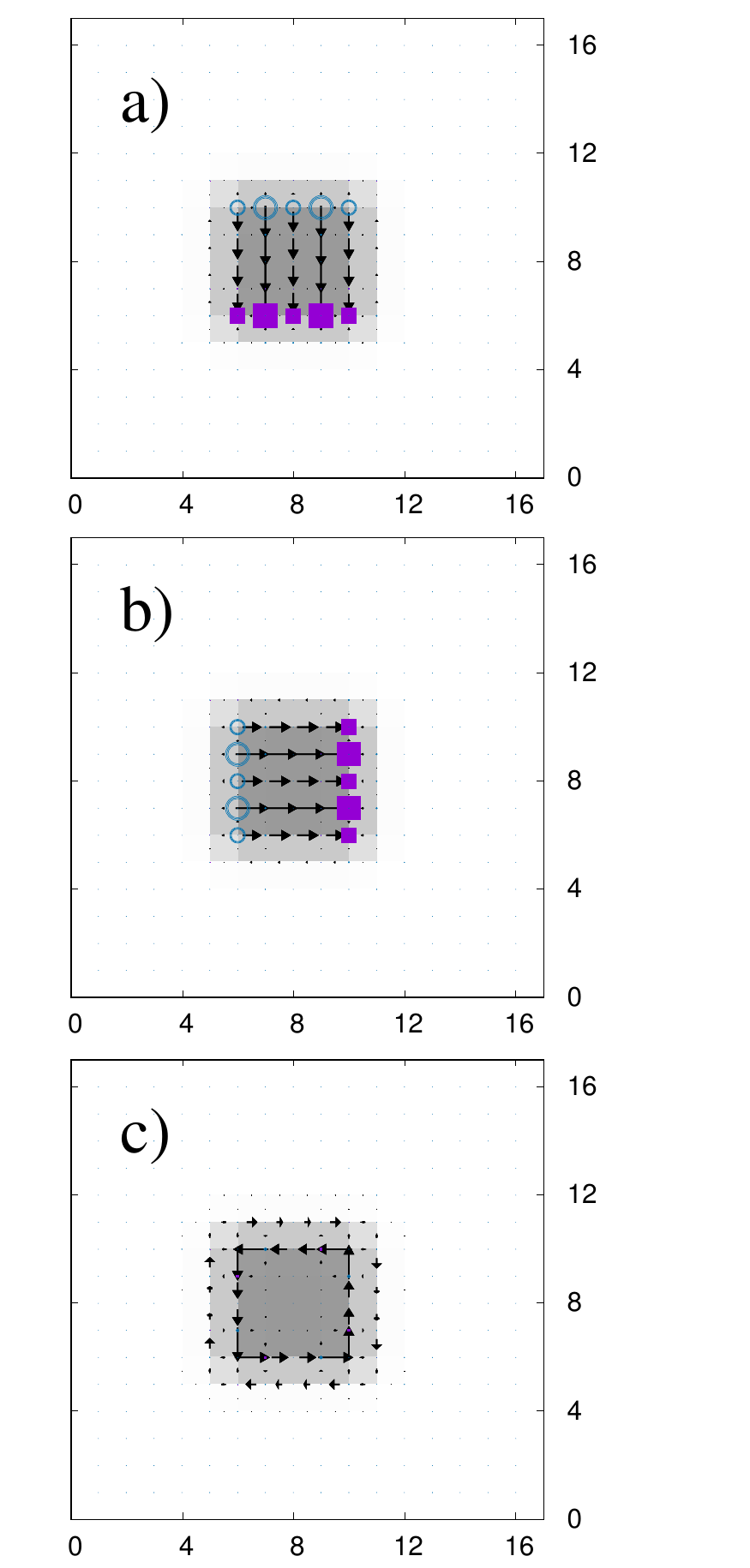}
\caption{(a-c) The $x$- (panel a), $y$- (panel b), and $z$- (panel c) component of spin currents
(arrows) and torques (circles, squares) for a phase separated
solution. The distribution of charge ($26$
particles on a $16 \times 16$ lattice) is indicated in grey. 
Parameters: $a_0=0.3$, $a_1=1.5$, $b_0=0$, $b_1=0$.}
\label{fig3}
\end{figure}
\begin{figure}[htb]
\includegraphics[width=7.5cm,clip=true]{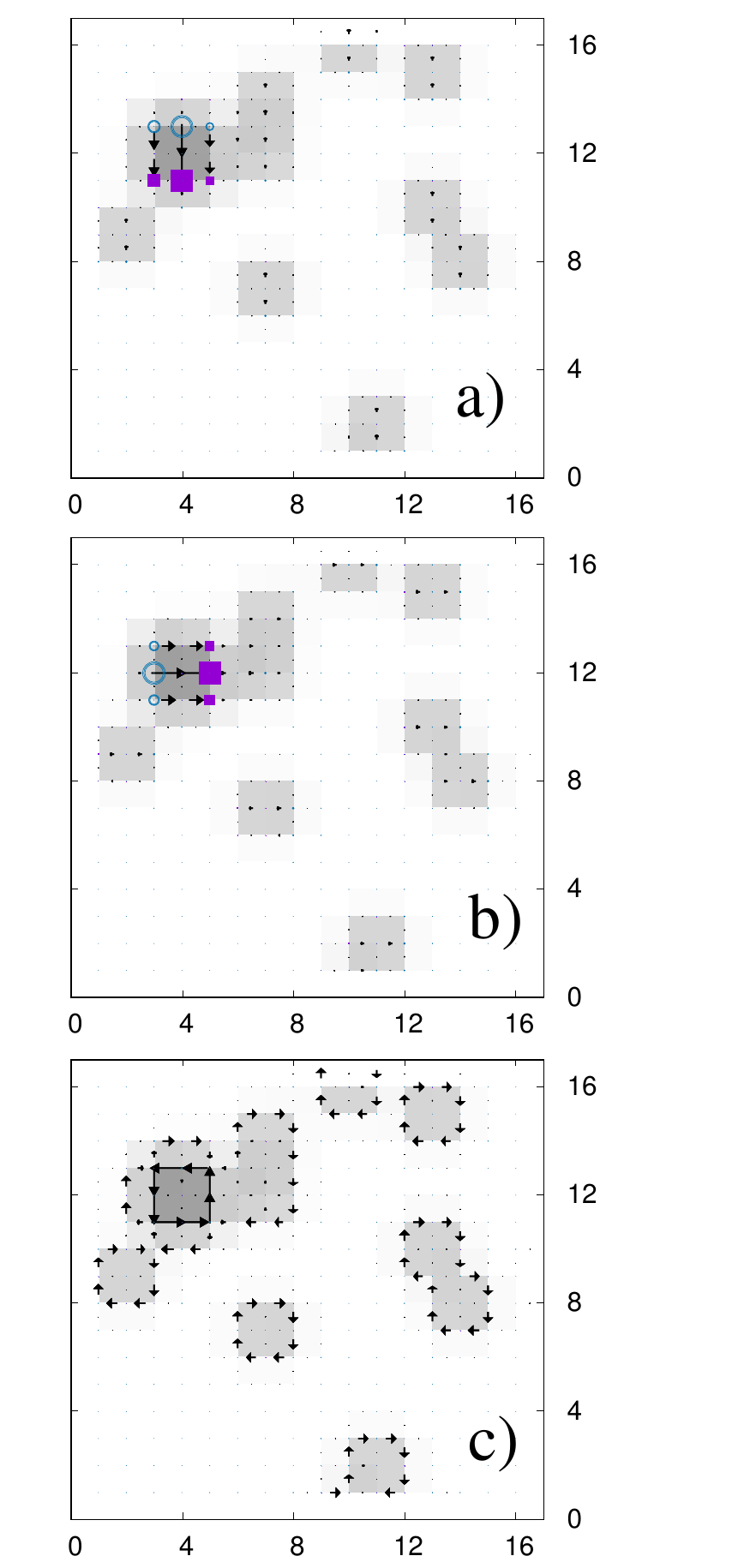}
\caption{(a-c) The $x$- (panel a), $y-$ (panel b), and $z-$ (panel c) component of spin currents
(arrows) and torques (circles, squares) for a phase separated
solution including disorder $V_0/t=0.5$. The distribution of charge ($26$
particles on a $16 \times 16$ lattice) is indicated in grey. 
Parameters: $a_0=0.3$, $a_1=1.5$, $b_0=0$, $b_1=0$.}
\label{fig4}
\end{figure}
Fig.\,\ref{fig3} reports a particular realization of a phase-separated solution obtained
on a $16\times 16$ lattice with $26$ particles. The charge carriers are confined to 
square shaped cluster with ``large density sites''  $n\approx 1$
and a border region with $n\approx 0.04$, indicated by dark and light grey squares, respectively.
As in the homogeneous case the dominant flow of the $x(y)$-spin currents
is along the $y(x)$ direction but now of course confined to the ``large density square''.
However, due to this confinement the currents are obviously not
conserved but finite torques lead to a generation (annihilation)
of spin currents.  This implies finite torques
$G^{x,y}$ for both $x$- and $y$- components at the border of the phase separated region
which in turn from Eq.\,(\ref{eq:gj}) induces $z$-polarized {\it edge} spin currents
flowing counter clockwise around the square.
Not that there is also a smaller edge current at the outer border (within the low
density region) flowing clockwise. This is due to small $x(y)$-spin currents in this region
(not visible on the scale of the plot) which flow opposite to the ones within the 
main square and thus are related to torques with opposite sign.

The pure phase separated state is very susceptible to the presence of disorder which
will break up the system into ``puddles'' with enhanced charge density.
This is shown in Fig.\,\ref{fig4} for a disorder strength $V_0/t=0.5$. 
$x$- and $y$- polarized spin currents are dominant in the extended puddle with large
charge density (around site $[4,12]$) but are also present (though not visible
on the scale of the plot) in the smaller puddles. 
Again the most interesting observation is the torque induced flow of $z$-polarized
spin edge currents around the puddles.


\section{Inhomogeneous Quantum Hall States}

\subsection{Momentum and real-space analysis of inhomogeneous QH states}
Since a sufficiently strong and density dependent RSOC can promote an inhomogeneous electron state at 
LXO/STO interfaces, it is worthwhile investigating the properties of this inhomogeneous electron gas under a strong magnetic field
$\mathbf B=B\hat{\bm z}$ perpendicular to the interface, in the quantum Hall regime. The Landau levels of a 2DEG in the presence of 
RSOC are \cite{rashba-qhe}
\[
E_s^\pm=\hbar\omega_c\left[s+\frac{1}{2}\pm\frac{1}{2}\mp\frac{\alpha}{\hbar}\sqrt{\frac{2m}{\hbar\omega_c}\left(s
+\frac{1}{2}\pm\frac{1}{2}\right)}\,\right]
\]
where $\omega_c=eB/m$ and we have taken the free-electron gyromagnetic factor $g=2$. As it is seen from the above equation,
the RSOC $\alpha$ lifts the degeneracy of the levels $E^+_s$ and $E^-_{s+1}$ even at $g=2$, so all levels have the same degeneracy 
as the ground state and host the same number of states $N_\phi$. Furthermore, the level spacing is not constant and in particular
the spacing between one level and the following with equal chirality decreases when the quantum number $s$ increases.
The ordering of the levels is not defined a priori: the level $E^+_{s+1}$ may fall below the level $E^-_{s + 1}$, provided the 
ratio $\alpha/\sqrt{B}$ is large enough. Only the level $E^-_{s=0}$ is independent of $\alpha$.

If the RSOC is constant, the chemical potential at $T=0$ is a non-decreasing step-wise function of the electron density, as
in the case $\alpha=0$. However, if the RSOC depends on the electron density, $\mu$ may decrease when jumping from one Landau
level to the next. Within the present continuum model we adopt a density dependent RSOC of the form 
\beq
\alpha(n)=\frac{2a_1n}{(1+2b_1n)^3}
\eeq
in agreement with Eq.\,(\ref{eq:coup}) once the identification $n_i=n_{i+\eta}=n$ is adopted in the continuum limit, and, 
for simplicity, $a_0=b_0=0$.
Furthermore, if the conditions required in Sec.\,\ref{rsoc-eps} are met, a situation like the one depicted 
in Fig.\,\ref{figure4-qhe} occurs, where the stepwise function $\mu(n)$ oscillates around the smooth curve $\mu_{B=0}(n)$, which
is itself a non-monotonic function of the density. Thus we may expect inhomogeneous quantum Hall states to occur.

\begin{figure}[h]
\includegraphics[width=7cm]{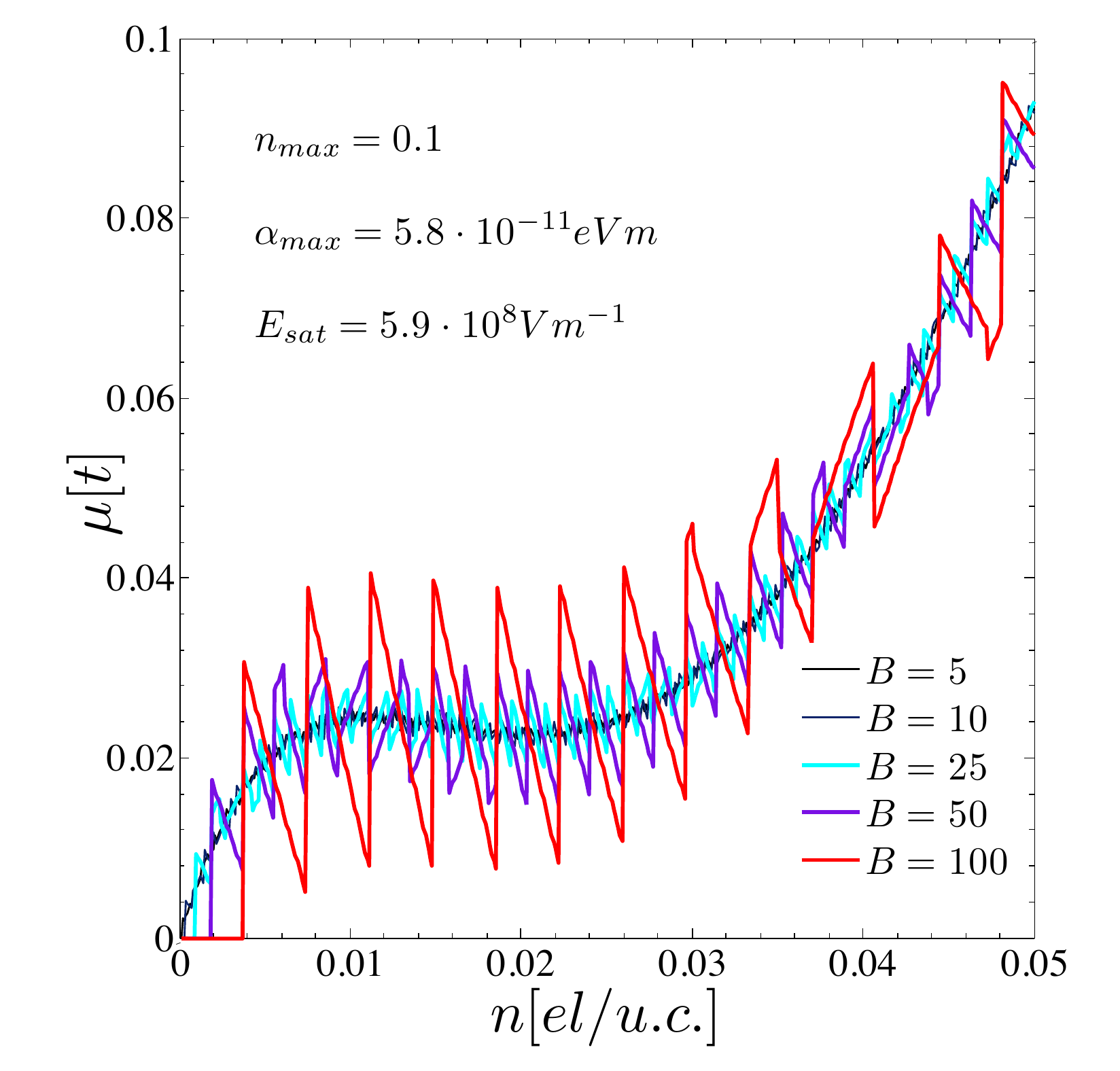}
\caption{Chemical potential as a function of the electron densities at various magnetic fields. The values of the 
parameters ($a_1$ and $b_1$) determining the RSOC are such that the maximum value $\alpha_{max}$ reported in the figure panel is reached 
at a density $n_{max}$ (per unit cell). The maximum value $E_{max}$ of the interfacial electric field is also reported. The units of the 
magnetic field $B$ are in Tesla.}
\label{figure4-qhe}
\end{figure}

To investigate the properties of inhomogeneous quantum Hall states, we performed calculations in real space, with the Hamiltonian
\begin{eqnarray*}
&&\mathcal H=\sum_{i,\sigma}\left[
t_{i,i+ x}\left(\mathrm e^{\mathrm i By_i}c^\dagger_{i+ x,\sigma}c_{i,\sigma}+h.c.\right)\right.\nonumber\\
&&\left.+t_{i,i+ y}\left(c^\dagger_{i+ y,\sigma}c_{i,\sigma}+h.c.\right)
+(B\sigma-\mu+\lambda_i)c^\dagger_{i,\sigma}c_{i,\sigma}\right]\nonumber\\
&&+\mathrm i \sum_{i,\sigma\sigma'}\left[\gamma_{i,i+ x}\left(\mathrm e^{-\mathrm i By_i}c^\dagger_{i,\sigma}
\tau^y_{\sigma\sigma'}c_{i+ x,\sigma'}+h.c.\right)\right.\nonumber\\
&&\left.-\gamma_{i,i+ y}\left(c^\dagger_{i,\sigma}
\tau^x_{\sigma\sigma'}c_{i+ y,\sigma'}+h.c.\right)\right]-\sum_i\lambda_i n_i,
\end{eqnarray*}
and the density dependent RSOC is described as before, by taking Eq.\,(\ref{eq:coup}) with $a_0=b_0=0$.
The results reported below are obtained for a square lattice of size $L=16$. In the absence of RSOC, the commensurability condition
requires that the magnetic flux through a unit cell $\phi_a$ is a rational fraction $p/q$ of the flux quantum $\phi_0$. 
Under this condition, the size of the magnetic unit cells $1 \times q$. Although for the chosen Landau gauge 
$\left[ {\bf A} = (By,0,0) \right]$ the vector potential breaks 
the translational invariance along $y$, the system preserves the symmetry for translations of $q$ lattice spacings along $y$
(see, e.g., Ref.\,\onlinecite{fradkin}). Then, 
if the system is composed by an integer number of magnetic cells along $y$, periodic boundary conditions (PBCs) can be imposed. In 
a $16 \times 16$ lattice, the lower field compatible with the latter condition corresponds to the ratio $p/q=1/16$, yielding
$B_{min}=1692$\,T, given the planar unit cell of LXO/STO. 
We notice in passing, that such a large unphysical magnetic field is only required to deal with  a small enough cluster to be numerically manageable.
Since the RSOC affects the QH states via the combination $\sqrt{ \alpha^2/B}$, the same physics can be obtained by choosing a
ten times smaller RSOC and a hundred times smaller field $B\sim 17$ T. This, however, gives rise to a ten times larger magnetic
length, that would require larger real-space clusters. In order to attack this problem with real-space calculations, we are therefore
led to use larger fields having in mind that the same physical effects would occur at much lower fields on
somewhat larger length scales. For the case at hand, the resulting Hofstadter spectrum is composed of 16 sub-bands, 
each of them accommodating 16 electrons. If spin is taken into account the number of the sub-bands doubles. If $N=16$ electrons 
are present, the ground state corresponds to the complete filling of the first level. The electron density is homogeneous and a 
current locally flows along $x$, within the chosen gauge (see Fig.\,\ref{figure5-qhe}).

\begin{figure}[h]
\includegraphics[width=6cm]{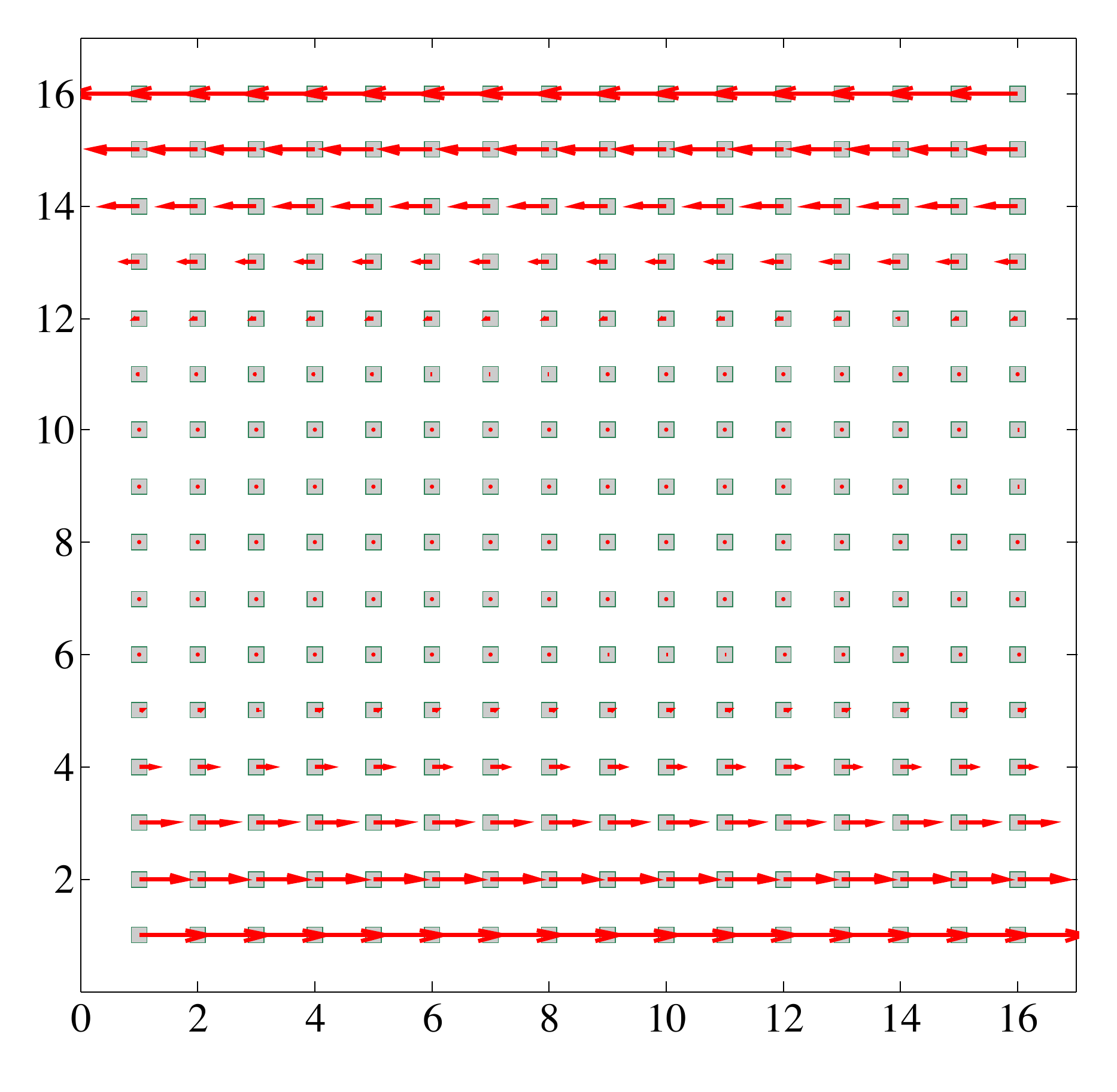}
\caption{Charge density distribution and charge current for a $16\times 16$ system with $N = 16$ electrons with PBCs. 
The lowest sub-band is completely filled.}
\label{figure5-qhe}
\end{figure}

The main features of quantum Hall states are deeply related to the existance of boundaries delimiting the physical space available 
for electron motion. When electrons are confined in a box, the wave functions must vanish approaching the walls. The effect of 
the boundaries is to lift the degeneracy of the Hofstadter sub-bands, that acquire a finite width. In other words, each sub-band in 
turn splits into a stack of levels. In Fig.\,\ref{figure6-qhe} for pedagogical reasons and for the 
sake of comparison, we show the charge distribution and the edge currents for systems with 
$N = 12$ (left) and $N = 16$ (right) electrons, respectively, with open boundary conditions (OBCs). In the first case, all 
the electrons are accommodated in the lowest 
sub-band and a single edge current goes through the sample. The charge density is substantially homogeneous in the bulk and decreases 
when approaching the boundaries. For $N = 16$, instead, the first sub-band is completely filled and the second one is partially 
filled, unlike the situation with PBCs where for $N = 16$ the latter was empty. The second sub-band is characterized by a negative 
conductance and two different edge states with currents flowing in opposite directions are achieved (left panel). If no spin-orbit 
coupling is present, the $z$-spin current is simply opposite to the charge current: when the electrons move towards the left, 
there is a net spin current along the direction of their motion, while the electric current is directed to the right.

\begin{figure}[h]
\includegraphics[width=8cm]{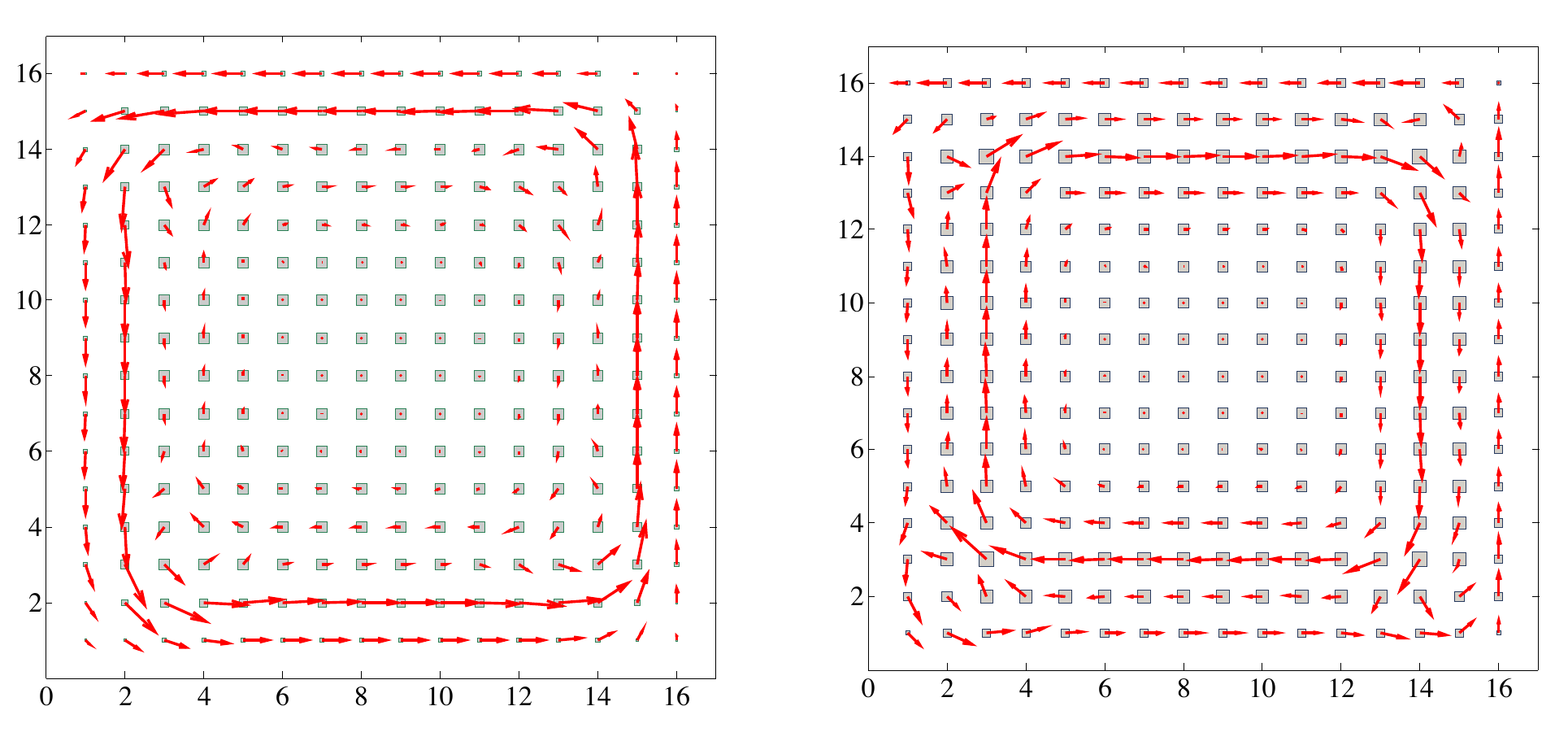}
\caption{Homogeneous states of a for a $16\times 16$ system in the quantum Hall regime without RSOC, for OBCs. 
Left: The lowest sub-band only is populated. Right: Opposite edge currents due to the filling of different Hofstadter 
sub-bands.}
\label{figure6-qhe}
\end{figure}

When RSOC (and its dependence on the local density) is taken into account, our real-space numerical analysis automatically carries 
out a minimization of the ($\lambda$ constrained) energy, by allowing inhomogeneous solutions when phase separation occurs. In 
Fig.\,\ref{figure7-qhe} we compare a situation in which the homogeneous system is inside the phase-separation region ($N=36$)
and a system in which the homogeneous system is outside the phase-separation region ($N=101$). As it is evident, in the former case, 
we obtain an inhomogeneous solution in our real-space calculation. Interestingly, the edge currents run along the boundary of
the self-nucleated droplet. 

\begin{figure}[h]
\includegraphics[width=8cm]{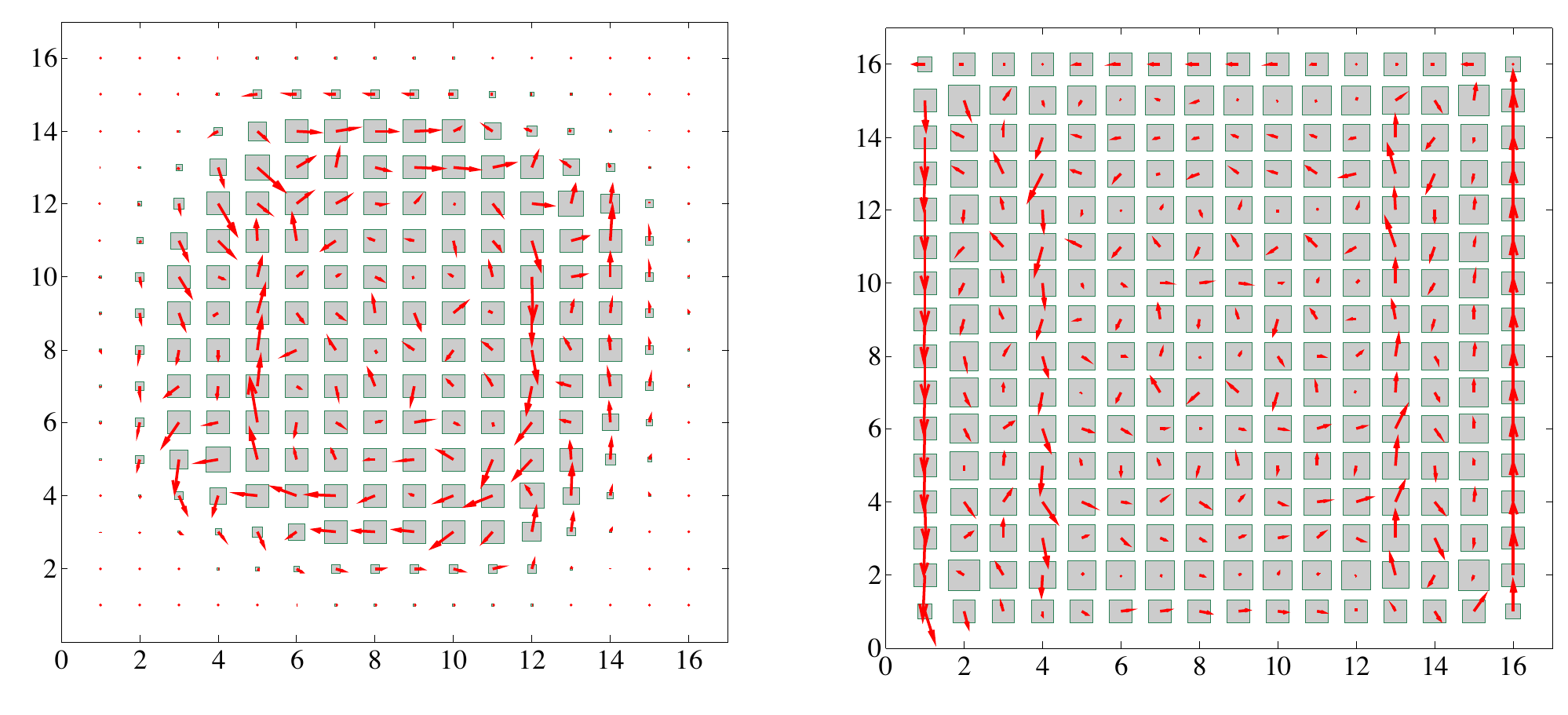}
\caption{Charge density distributions and charge currents for a $16\times 16$ system. Left: Inhomogeneous quantum Hall state 
for a number of electrons 
$N = 36$, corresponding to a filling at which
the infinite homogeneous systems falls inside the phase-separation region. Right: Homogeneous quantum Hall state for a number 
of electrons $N = 101$, corresponding to a filling at which the infinite homogeneous systems falls outside the 
phase-separation region.}
\label{figure7-qhe}
\end{figure}

\subsection{Lattice model for QHE: Harper equation with RSOC}
We consider a square lattice infinite in the $x$-direction and extended over $L$ unit cells 
(lattice constant $a=1$) in the $y$-direction. 
Electrons can hop between neighboring sites in the $xy$-plane and are subject to a strong homogeneous magnetic field 
$\mathbf B = B \hat{\bm z}$ generated by a site-dependent vector potential $\bm A (j) =(-By_j,0,0)$.
The tight-binding Hamiltonian in presence of RSOC is 
$H_{\rm TB} = H_{0} + H_{\rm RSOC}$ with
\beqa
H_{0}&=&-\sum_j t_{j,j+x} \Big[e^{i 2\pi \phi y_j}c^{\dagger}_{j+x}c_j + h.c. \Big ]  \\
&-& \sum_j t_{j,j+y} \Big [ c^{\dagger}_{j+y}c_j + h.c. \Big ]- E_{\rm Z} \sum_{j \sigma \sigma'} 
\Big [ c^{\dagger}_{j \sigma} \tau^z_{\sigma \sigma'}
c_{j \sigma'} \Big ] ,  \nonumber 
\eeqa
\beqa
H_{\rm RSOC}&=&\mathrm
 i \sum_{j,\sigma,\sigma'}\gamma_{j,j+x}\Big [ e^{i 2\pi \phi y_j}c^{\dagger}_{j,\sigma}\tau^y_{\sigma \sigma'}c_{j+x,\sigma'}
 +h.c \Big ]
\nonumber \\
&-&\mathrm i\sum_{j,\sigma,\sigma'}\gamma_{j,j+y} \Big [ c^{\dagger}_{j,\sigma}\tau^x_{\sigma \sigma'}c_{j+y,\sigma'}+h.c \Big ] .
\eeqa
The orbital effect of the magnetic field is encoded in the phase factor acquired by the hopping amplitudes

\[
t_{j,j+x(y)} \longrightarrow t_{j,j+x(y)} e^{\mathrm i e/\hbar  \int_{j}^{j+x(y)}{\mathbf{A}\cdot \mathbf{dr}}},\,\,\,\,\,\,\, 
t_{j,j+x(y)} = t_{x(y)}
\]
and 
\[
\gamma_{j,j+x(y)} \longrightarrow \gamma_{j,j+x(y)} e^{\mathrm i e/\hbar  
\int_{j}^{j+\hat{x}(\hat{y})}{\mathbf{A}\cdot \mathbf{dr}}},\,\,\,\,\,\, \gamma_{j,j+x(y)} = \gamma_{x(y)}
\]
which can be written in terms of the flux $\phi$ through a lattice cell (in units of the flux quantum $\phi_0 = h/e$). 
$\tau^x$, $\tau^y$ and $\tau^z$ are Pauli matrices acting on the electron spin and $E_{\rm Z}=g2\pi \phi $ the Zeeman coupling constant. 
An eigenstate of $H_{\rm TB}$ can be expanded as 
$|\Psi\rangle=\sum_{j \sigma} \psi_{\sigma}(x_j,y_j)c^{\dagger}_{j \sigma}|0\rangle$, with the 
wavefunction $\psi_{\sigma}(x_j,y_j)=\psi_{\sigma}(\ell,m)$ centered on the lattice site of coordinates $x_j=\ell$ and $y_j=m$. 
Translational invariance along $x$ allows for the factorization $\psi_{\sigma}(\ell,m)=\psi_{\sigma}(m) e^{i k_x \ell}$ and the
eigenvalue problem can be solved in a ribbon of vertical size $L$. 

\begin{figure}[htb]
\includegraphics[width=5.2cm]{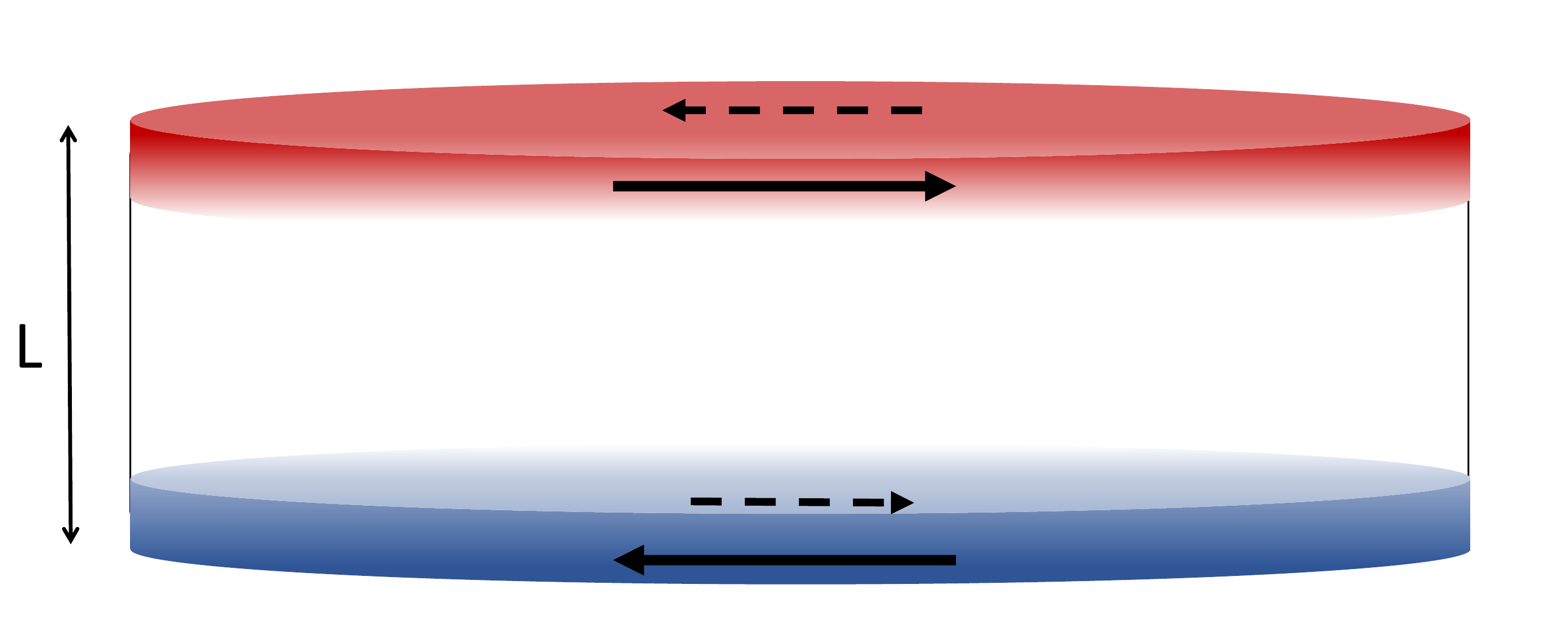}
\caption{Schematic view of a ribbon with periodic boundary conditions along $x$ and finite size $L$ along $y$.
For an homogeneous state only two edge states are present: the upper edge state (red) flows counterclockwise, while the lower 
edge (blue) flows clockwise.}
\label{ribbon}    
\end{figure}

The Schr\"odinger equation for the spinor 
\begin{equation}
\Psi_{m}=
\left (
\begin{aligned}
&\psi_{m \uparrow}\\
&\psi_{m \downarrow}
\end{aligned}
\right )\,\,\,\,\, m=0,\cdots,L-1
\end{equation}
reads as
\begin{equation}
E \Psi_m = D_m \Psi_m + R_m \Psi_{m+1} + R_m^{\dagger} \Psi_{m-1},
\label{harper_eqs}
\end{equation}
known as Harper equation \cite{hofstadter-1976,demikhovskii-2006,morais-smith-2012}.
{\rm RSOC} enters in the off-diagonal elements of the $2\times 2$ blocks
\begin{equation}
\begin{aligned}
D_m &=
\begin{pmatrix}
-2t_x \cos (\tilde{k}_x ) - E_{\rm Z} & 2 \mathrm i \gamma_x \sin (\tilde{k}_x)\\
-2 \mathrm i \gamma_x \sin (\tilde{k}_x) &-2t_x \cos (\tilde{k}_x ) + E_{\rm Z}
\end{pmatrix} \nonumber \\ 
R_m &=
\begin{pmatrix}
-t_y &\ -\mathrm i \gamma_y\\
-\mathrm i \gamma_y &-t_y\\
\end{pmatrix},
\end{aligned}
\end{equation}
where $\tilde{k}_x\equiv k_x+2\pi \phi m$.
The solution of the coupled eigenvalue equations defined by Eq.\,(\ref{harper_eqs}) returns $2L$ energy sub-bands 
$E_\ell(k_x)$ with $\ell = 0, 1, \cdots, 2L-1$. 
It is convenient to take the origin of the $y$-axis $m=0$ at the middle of the ribbon, so that $m$ takes the integer values 
between $-L/2$ and $L/2-1$ (for simplicity we assume $L$ to be even here). 
\begin{figure}[htb]
\includegraphics[width=8cm]{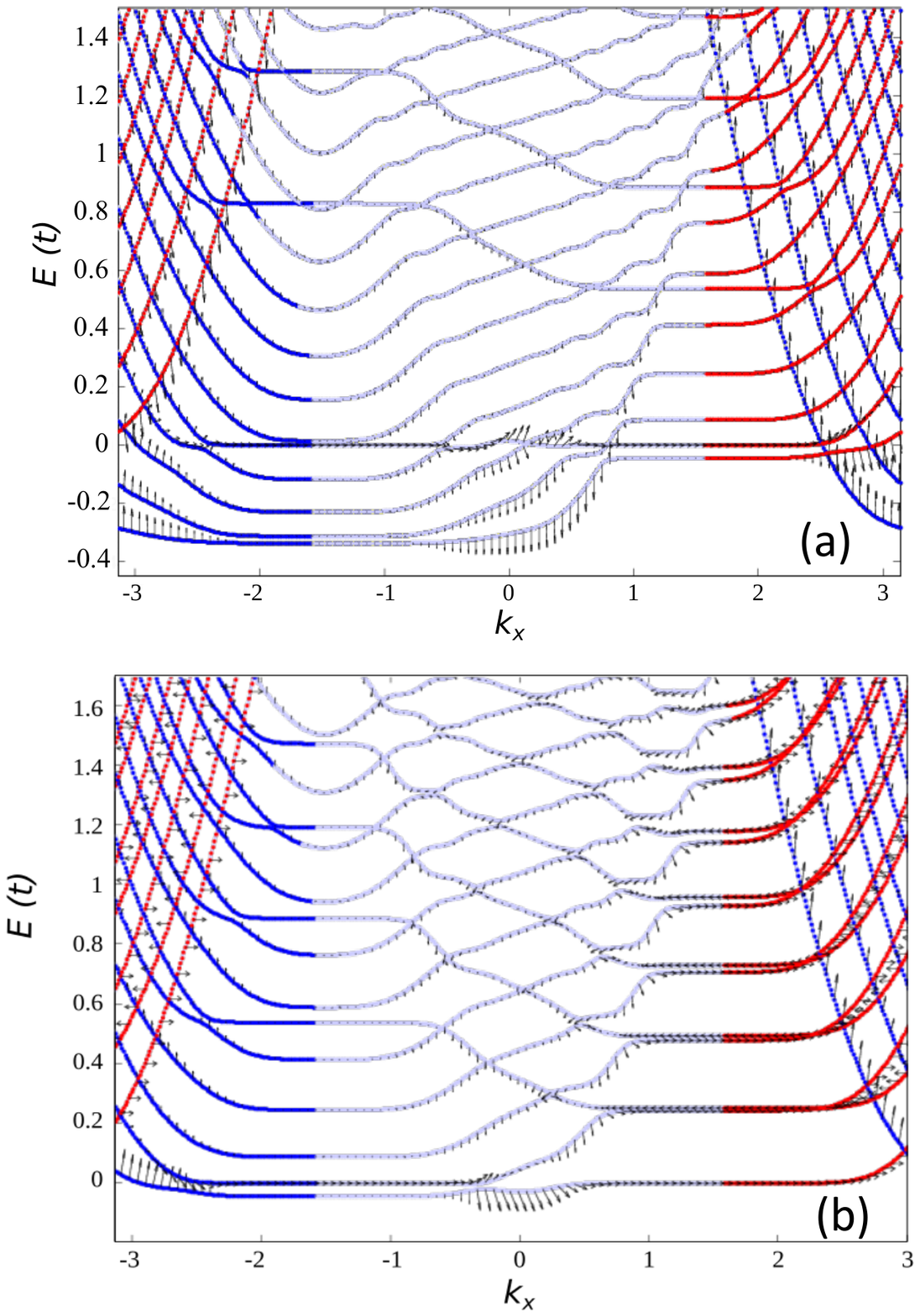}
\caption{Electronic sub-bands for the two-dimensional system in strong magnetic field for (a) 
 heterostructure formed by a region with {\rm RSOC} ($\gamma_1=a_{01}=0.3t$ and a 
region with {\rm RSOC} $\gamma_2 = a_{02}=0.6t$;
(b) heterostructure formed by a region without {\rm RSOC} ($\gamma_1=0$ and a 
region with {\rm RSOC} $\gamma_2 = a_0=0.3t$. Colors distinguish the edge states --
propagating along the top edge from left to right (in red) and along the bottom edge from right to left (in blue) -- 
from the bulk states (light grey), 
according to the calculated expectation value of the $y$-coordinate $\langle m \rangle=\sum_{m \sigma} m |\psi_{m \sigma}|^2$. 
Black arrows represent magnitude and direction of the expectation value of the spin angular-momentum: vertical arrows stay 
for $\langle \sigma_y \rangle=1,\,\langle \sigma_z \rangle=0$, horizontal arrows for 
$\langle \sigma_y \rangle=0,\,\langle \sigma_z \rangle=1$.}
\label{ens_kx}    
\end{figure}
In order to investigate inhomogeneous quantum Hall states we consider an interface in the $y$-direction between two 
(macroscopic) regions with different RSOC $\gamma_1$ and $\gamma_2$ (which might result from different electronic densities 
in the two regions due to a density-dependence of the RSOC) and compare the spectrum -- as a function of the momentum $k_x$ 
-- for this heterostructure with the conventional spectrum of Landau levels in absence of {\rm RSOC} and with the case where 
RSOC is present but is homogeneous. Given a ribbon as in Fig.\,\ref{ribbon}, numerical results are 
shown in Fig.\,\ref{ens_kx} for the different cases: { (a)} $\gamma_{y(x)} = \gamma_1=0.3t$ for $0 < m \leq L/2-1$, 
$\gamma_{y(x)} = \gamma_2=0.6t$ for $ 0 > m \geq -L/2$ and $\gamma_y = (\gamma_1 + \gamma_2)/2$ at $m=0$;
{(b)} $\gamma_{y(x)} = \gamma_1=0$ for $0 < m \leq L/2-1$, 
$\gamma_{y(x)} = \gamma_2=0.3t$ for $ 0 > m \geq -L/2$ and $\gamma_y = (\gamma_1 + \gamma_2)/2$ at $m=0$.

At $\phi \ll 1$ the bulk spectrum consists of a set of flat bands (Landau levels) which have different ordering 
and spin-polarizations depending on whether RSOC is present or not. 
Of course, inhomogeneous QH states have been investigated before (see, e.g., Ref.\,\onlinecite{venturelli-2011,champel-2013,hashimoto-2008,lado-2013} 
and references  therein). However, it is interesting that we face here an inhomogeneous QH state where different strengths of the local RSOC 
induce differently spin-polarized edge states. 
In Fig.\,\ref{ens_kx}(a,b) one can distinguish two sets of bulk levels which are connected 
at $k_x \approx 0$. It is interesting to note the avoided crossings between levels with different quantum numbers and the variation 
of the orientation of the spin particularly along the lowest energy levels. On the one hand, case (a) is rather similar to 
the case of no RSOC and differently gated regions of the system that was considered in Ref. \onlinecite{venturelli-2011}. The main difference
here is that the presence of a sizable RSOC forces the spin polarization of the edge states (moving in the $x$ direction) along the $y$ direction
instead of the usual $z$ direction. On the other hand, in the case of Fig.\,\ref{ens_kx}(b) the upper  edge lives 
in a region of vanishing RSOC and is  polarized along $z$, while the blue  (i.e. lower) edge lives in a region of sizable RSOC and carries a 
chiral spin polarized along $y$. The corresponding edge states that mix and interfere inside the bulk of the ribbon give rise to smoothly
rotating spin polarizations. These  effects, might be of applicative relevance for spin interferometry.\cite{giovannetti-2012}
or they  might play important roles in electronic  
transport. Note that even at $\gamma \neq 0$ there is always a level with electrons having the spin polarized in the $z$-direction, 
regardless of the momentum $k_x$ (the $s=0$ level in the previous section).

\section{Spin Hall effect}
In the ground state of the Rashba model the total $x(y)$-torques have to vanish and therefore from Eq.\,(\ref{eq:gj}) we get
\begin{equation}\label{eq:gj2}
\sum_i G_i^{x(y)}=0=-\gamma_{i,i+x(y)} J^z_{i,i+x(y)} - \gamma_{i-x(y),i} J^z_{i-x(y),i},
\end{equation}
which implies also a vanishing of the total $z$-polarized
spin currents for $\gamma_{i,i+x(y)}=const$. 
However, consider for example a system with striped RSOC as depicted
in the inset to Fig.\,\ref{fig5}b.
Denote with $J^z_{1,2}$ the total z-spin current
flowing along the bonds of the $\gamma_{1,2}$-stripes which are assumed to have
the same width.
Then we can rewrite Eq.\,(\ref{eq:gj2}) as
\[
0=\gamma_1 J^{z}_1 + \gamma_2 J^z_2 \,\,\longrightarrow\,\, J^z_2=-\frac{a_1}{a_2}J^z_1 
\]
and the total $z$-spin current of the system is thus given by
\[
J^z_{tot}=n_{str}\left(J^{z}_1+J^{z}_2\right)=n_{str}J^{z}_1\left(1-\frac{\gamma_1}{\gamma_2}\right) \,,
\]
where $n_{str}$ denotes the total number of $\gamma_{1,2}$ stripes.

\begin{figure}[htb]
\includegraphics[width=7.5cm,clip=true]{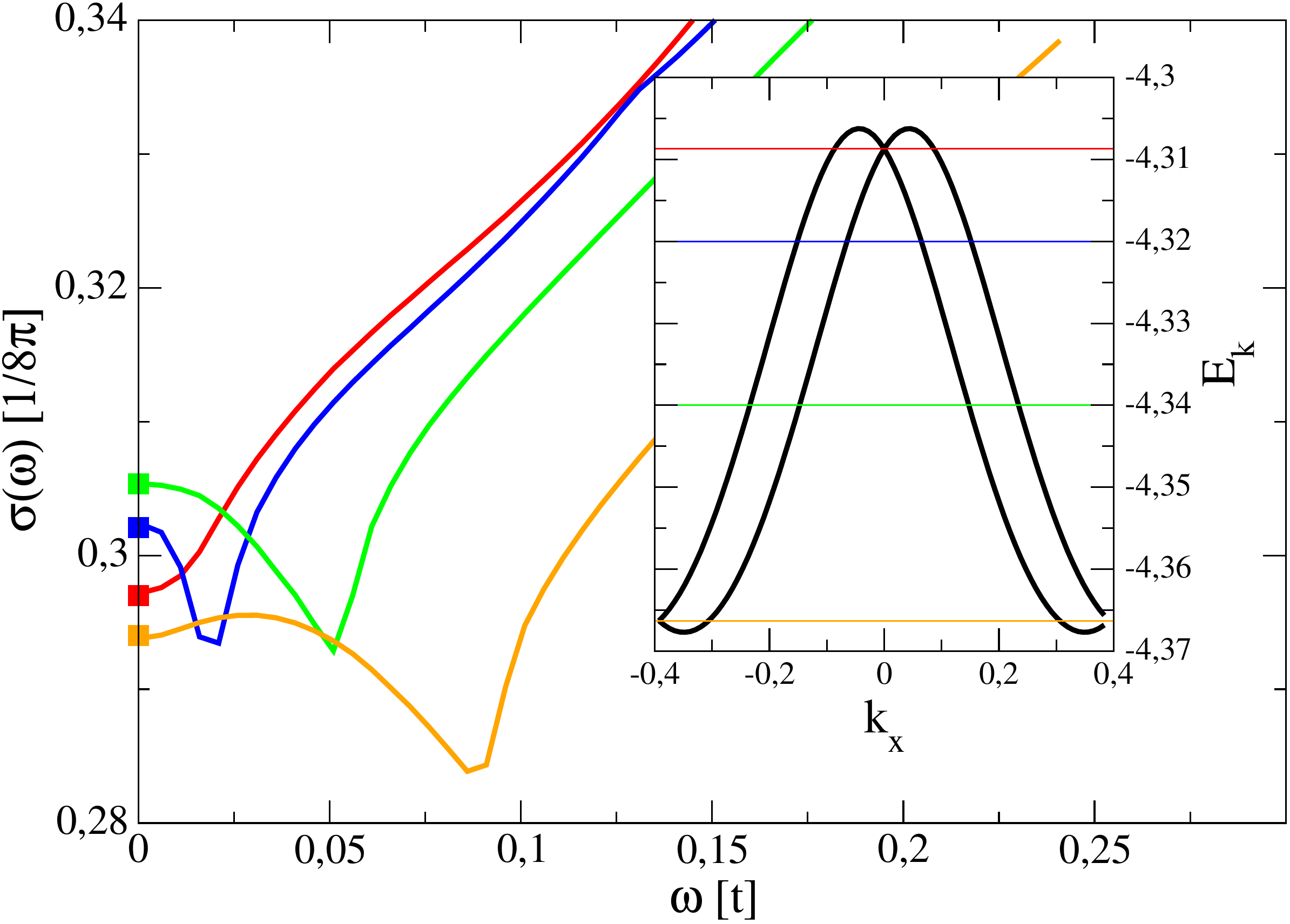}
\includegraphics[width=7.5cm,clip=true]{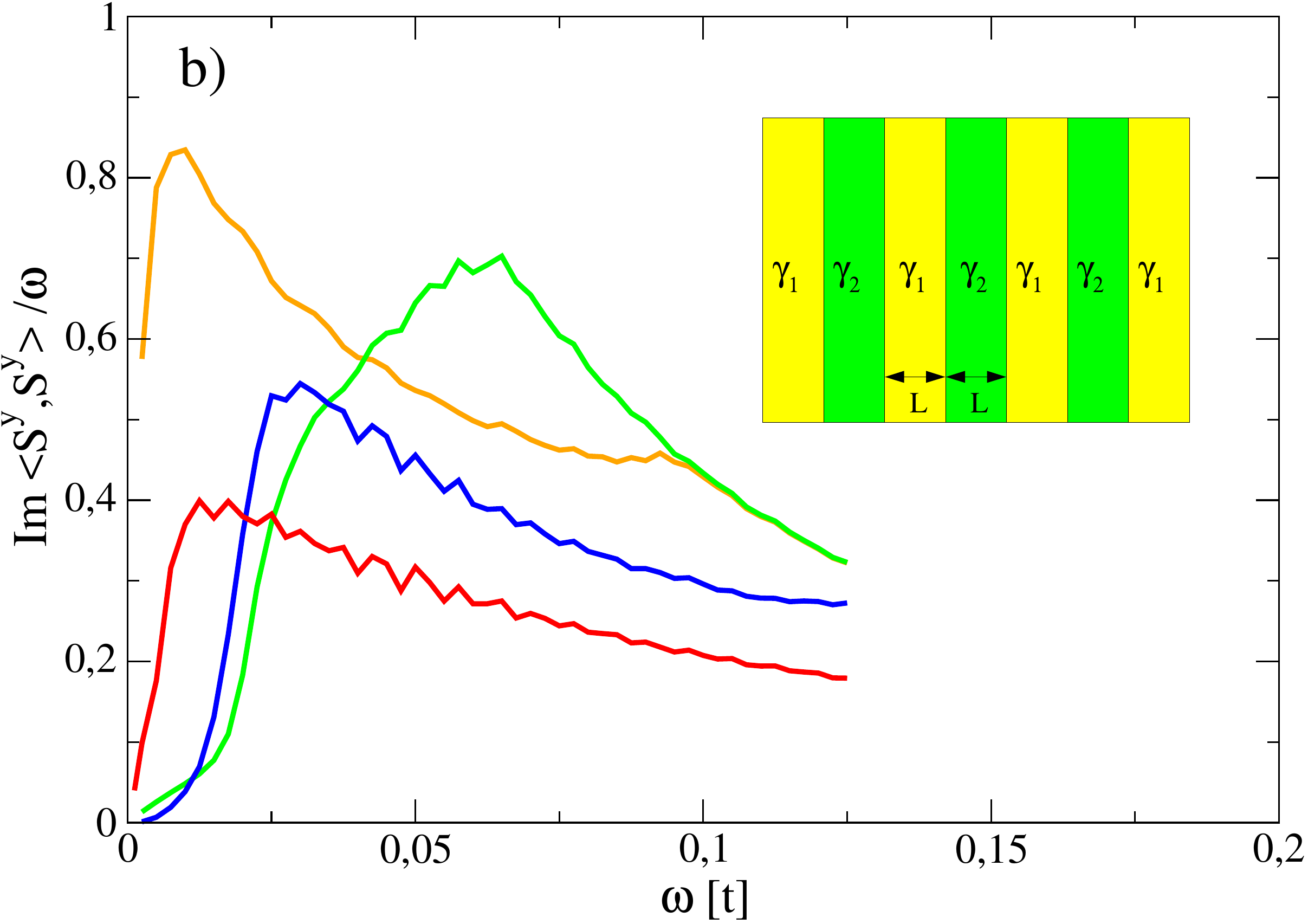}
\caption{Top panel: Frequency dependence of the spin Hall correlation function
Eq.\,(\ref{eq:kubo1}) for different values 
of the chemical potential which is located within the lowest bands as indicated
in the inset.
Lower panel: Imaginary part of the response related to the time derivative
of $S^y$. The inset depicts the coupling structure of the striped
Rashba system ($L=4$) with RSOC $\gamma_1=0.2t$ and $\gamma_1=0.8t$, respectively.}
\label{fig5}
\end{figure}

The same reasoning can also be applied in a non-equilibrium situation, i.e.
in the presence of an applied electric field, where it gives rise
to the so-called spin Hall effect (SHE) \cite{dyakonov1971}, i.e. 
the generation of a transverse
spin current by an applied electric field with the current
spin polarization being perpendicular to both the field
and the current flow.
For a homogeneous system (with homogeneous linear RSOC)  
the SHE vanishes in a stationary situation because of the same argument, which leads to Eq.\,(\ref{eq:gj2})
and which has been first pointed out by Dimitrova \cite{dimitrova05}. 
On the other hand we have shown in Refs. \onlinecite{epl,jscmag, jmagmat} that in the linear
response regime a periodic modulation of the RSOC as depicted in Fig.\,\ref{fig5}
generates a finite SHE within the low doping doping regime where the electronic
states are localized but can sustain a finite spin current under stationary conditions.

The spin Hall coefficient can be obtained as the zero frequency limit
of the following spin-charge current correlation function as
\begin{eqnarray}
\sigma^{sH}(\omega)&=&\sum_{ij}\Re\sigma^{reg}_{ij}(\omega)\label{eq:kubo1}\\
\sigma^{reg}_{ij}(\omega)&=& -\frac{\mathrm i e}{N}\sum_{\ell,m}\frac{f_\ell-f_m}{E_\ell-E_m}
\frac{\langle \ell|j^z_{i,i+y}|m\rangle
\langle m|j^{ch}_{j,j+x}|\ell\rangle}{\omega+\mathrm i\eta+E_\ell-E_m}, \, \nonumber
\end{eqnarray}
 where $|\ell\rangle$ and $E_\ell$ are exact eigenstates and eigenvalues
  of the system and $f_\ell$ denotes the corresponding Fermi function. Due to the
  modulation of the RSOC, there a band folding producing a sub-band structure in
  the reduced Brillouin zone. 
Fig.\,\ref{fig5}a shows the frequency dependence of $\sigma^{sH}(\omega)$
for a $L=4$ striped system (cf. inset to panel b) and a set of chemical
potentials within the lowest pair of Rashba split sub-bands.
Clearly $\sigma^{sH}(\omega)$ approaches a finite value for $\omega\to 0$
which is slightly reduced when the chemical potential corresponds
to the energy of a band crossing.
However, one has to additionally show that this is a result
obtained under stationary conditions. This can be substantiated
from the computation of $\Im \chi(S^y,S^y)/\omega$ which corresponds
to the dynamics of the time derivative of $S^y$.
From Fig.\,\ref{fig5}b it turns out this quantity vanishes in the
limit of $\omega\to 0$ due to the strong localization of the bands
perpendicular to the stripe direction. The vanishing of $dS^y/dt$
can also be derived directly from the equation of motion for $S^y$ which
then provides another validation for Eq.\,(\ref{eq:gj2}) \cite{epl}.

It is important to note that this effect is due to the modulation of the RSOC
and cannot arise in a conventional charge-density wave system.
Experimentally our proposal could be realized in the 2DEG at the interface of
a LaAlO$_3$/SrTiO$_3$ ( LAO/STO)  heterostructure with periodic top gating
or in heterostructures of semiconductors with 
modulated Rashba SOC which have been already discussed
in the literature in different contexts \cite{wang04,japa09,malard11}.

\section{Conclusions}
In this paper we discuss the role of density inhomogeneities in oxide interfaces. These inhomogeneities seem to be
a common (if not ubiquitous) occurrence in LXO/STO oxide interfaces and likely have a submicrometric character. 
The inhomogeneous density distribution is directly related to an inhomogeneous distribution of the
confining electric field perpendicular to the interface and, in turn, to an inhomogeneous distribution of
the RSOC. This naturally leads to the idea that the RSOC can be phenomenologically assumed as a function
of the local density, that increases at low density and is implemented in the simplified model presented in Sect. III.  The RPA analysis 
and the real-space solution of the model on finite clusters show that a phase-separation instability is present, which also preemps 
the occurrence of an electronic instability at finite momentum. 

The explicit solution of the model in real space finite lattices directly shows that the density-driven RSOC can indeed 
induce (or at least makes it easier) an EPS and that the high-density phase is fragmented into `puddles' when disorder is present. 
This inhomogeneous distribution of electrons and RSOC gives rise to spin torques at the puddle boundaries. 
These torques act as sources and drains of spin currents, which tend to flow inside and around the puddles. 

Spin and charge currents also flow at the boundaries of the density inhomogeneities, when strong magnetic fields drive the system into a quantum Hall state.
 When the edge is between a metallic filled region and an empty insulating  matrix, the edge state assumes a simpler structure with the RSOC linking the spin
to the momentum inducing a chiral edge state with the spin polarization on the interface plane.  The edge states are instead more intricate and
display a complex behavior at the interface between regions with differently filled Landau levels. 

In summary, electronic inhomogeneities in oxide interfaces can induce inhomogeneous RSOC leading to interesting new physical effects (inhomogeneous 
spin currents and edge states). Of course these effects can be exploited and engineered for applicative purposes. This paves the
way to effects of interest for spintronics, like spin interferometry or the occurrence of a robust spin Hall effect in systems with striped RSOC.

{\bf Acknowledgments}

The Authors gratefully thank L. Benfatto and  C. Castellani for stimulating discussions. This work was financially supported by the
Ateneo 2016 project `Superconductivity in soft electronic matter'  of the University of Rome `Sapienza'


\begin{thebibliography}{10}
\bibitem{ohtomo-2004} A. Ohtomo and H. Y. Hwang, (2004) Nature {\bf 427}, 6 

\bibitem{reyren-2007} N.  Reyren, {\it et al.}, (2007) Science {\bf 317},  1196 

\bibitem{caviglia-2008} A. Caviglia, {\it et al.}, (2008) Nature {\bf 456},  624

\bibitem{biscaras-2010} J.  Biscaras {\it et al.}, (2010) Nat. Commun. {\bf 1}, 89 

\bibitem{biscaras-2012} J. Biscaras, {\it et al.},   (2012) Phys. Rev. Lett. {\bf 108},  247004

\bibitem{ariando-2010} Ariando, {\it et al.}, (2011) Nat. Commun. {\bf 2},  188

\bibitem{lilu-2011} Li Lu, C. Richter, J. Mannhart, and R. C. Ashoori, (2011) Nat. Phys. {\bf 7},  762 


\bibitem{bert-2011} J. A. Bert {\it et al.},  (2011) Nat. Phys. {\bf 7}, 767

\bibitem{dikin-2011} D. A. Dikin, {\it et al.}, (2011) Phys. Rev. Lett. {\bf 107}, 056802

\bibitem{metha-2012} M. M. Mehta, {\it et al.}, (2012) Nat. Commun. {\bf 3},  955

\bibitem{bert-2012} J. A. Bert , {\it et al.}, (2012) Phys. Rev. B {\bf 86}, 060503(R) 

\bibitem{Prawiroatmodjo-2016} G. E. D. K. Prawiroatmodjo, {\it et al.} (2016), Phys. Rev. B {\bf 93} 184504.

\bibitem{caviglia-2012} A. D. Caviglia, {\it et al.}, (2010) Phys. Rev. Lett. {\bf 104}, 126803

\bibitem{ben-shalom-2010} M. Ben Shalom,  {\it et al.},  (2010) Phys. Rev. Lett. {\bf 105}, 206401.

\bibitem{caprara-2012}S.  Caprara, F. Peronaci, and M. Grilli, (2012), Phys. Rev. Lett. {\bf 109}, 196401

\bibitem{hurand-2015} S. Hurand, {\it et al.}, (2015), Sci. Rep.  {\bf 5} 12751.

\bibitem{rashba-1984} Y. A. Bychkov  and E. I. Rashba, (1984), J. Phys. C {\bf 17} 6039 

\bibitem{bell-2009} C. Bell, {\it et al.} (2009), Phys. Rev. Lett. {\bf 103}, 226802

\bibitem{caprara-2011} S. Caprara, M. Grilli, L. Benfatto,  and C. Castellani (2011), Phys. Rev. B {\bf  84}, 014514

\bibitem{bucheli-2013} D. Bucheli, S. Caprara, C. Castellani, and M. Grilli (2013), New J. Phys. {\bf 15} 023014

\bibitem{ristic-2011} Z. Ristic Z, {\it et al.} (2011) Europhys. Lett. {\bf 93}, 17004

\bibitem{feng-bi-2013} Feng Bi {\it et al.} (2013) arXiv:1302:0204

\bibitem{bucheli-2015} D. Bucheli, S. Caprara, and M. Grilli, (2015) Supercond. Sci. Technol. {\bf 28}, 045004 

\bibitem{richter-2013}  C. Richter {\it et al.} (2013) Nature {\bf 502} 528

\bibitem{kalisky-2013} Beena Kalisky, {\it et al.}, (2013) Nat. Mat. {\bf 12} 1091

\bibitem{honig-2013} M. Honig, J. A. Sulpizio, J. Drori, A. Joshua, E. Zeldov and S. Ilani, (2013) Nat. Mat.  {\bf  12} 1112

\bibitem{caprara-2013} S. Caprara {\it et al.} (2013) Phys. Rev. B {\bf 88} 020504(R).

\bibitem{caprara-2015} S. Caprara, {\it et al.}, (2015) Supercond. Sci. Technol. {\bf 28} 014002 

\bibitem{biscaras-2013} J. Biscaras {\it et al.},  (2013) Nat. Mater. {\bf 12} 542

\bibitem{nanobridges} D. Stornaiuolo,{\it et al.}, (2012) App. Phys. Lett. {\bf 101}, 222601 

\bibitem{treske-2015} Uwe Treske, Nadine Heming, Martin Knupfer, Bernd B\"uchner, Emiliano Di Gennaro, Amit Khare, Umberto Scotti Di Uccio, Fabio Miletto Granozio, Stefan Krause, and Andreas Koitzsch, Universal electronic structure of polar oxide hetero-interfaces, Sci. Rep. in press.

\bibitem{salluzzo-2009} M. Salluzzo, J. C. Cezar, N. B. Brookes, V. Bisogni, G. M. De Luca, C. Richter, S. Thiel, J. Mannhart, M. Huijben, A. Brinkman, G. Rijnders, and G. Ghiringhelli, (2009) Phys. Rev. Lett. {\bf 102}, 166804.

\bibitem{zhong-2013} Z. Zhong, A. T\"oth, and K. Held, (2013) Phys. Rev. B {\bf 87}, 161102(R).

\bibitem{popovic-2008} Z. Popovi\`c, S. Satpathy, and R. Martin, (2008) Phys. Rev. Lett. {\bf 101}, 256801

\bibitem{delugas-2011} P. Delugas, A. Filippetti, V. Fiorentini, D. I. Bilc, D. Fontaine, and Ph. Ghosez (2011) Phys. Rev. Lett. {\bf 106} 166807

\bibitem{nakagawa-2006} N. Nakagawa, H. Y. Hwang, and D. A. Muller, (2006) Nat. Mater. {\bf 5} 204.

\bibitem{liping-yu-2014}Liping Yu and Alex Zunger,  (2014) Nat. Commun. {\bf 5}, 5118

\bibitem{bucheli-2014} D. Bucheli, M. Grilli, F. Peronaci, G. Seibold, and S.
Caprara, (2014) Phys. Rev. B {\bf 89}, 195448 

\bibitem{scopigno-2016} N. Scopigno {\it et al.},  (2016) Phys. Rev. Lett. {\bf 116}, 026804

\bibitem{fradkin} E. Fradkin, {\it Field Theories of Condensed Matter systems}, Addison-Wesley Publishing Company, 1991.

\bibitem{dyakonov1971}  M. I. Dyakonov and V. I. Perel, (1971) Phys. Lett. A {\bf 35}, 459 

\bibitem{epl15ps} G. Seibold, D. Bucheli, S. Caprara, and M. Grilli,  (2015) EPL {\bf 109}, 17006.

\bibitem{dimitrova05} Ol'ga V. Dimitrova, (2005) Phys. Rev. B {\bf 71}, 245327.

\bibitem{epl} G. Seibold, S. Caprara, M. Grilli, and R. Raimondi,  (2015) EPL {\bf 112}, 17004.

\bibitem{jscmag} G. Seibold, S. Caprara, M. Grilli, and R. Raimondi, (2017) Journal of Superconductivity
	 and Novel Magnetism {\bf 30}, 123.

\bibitem{jmagmat} G. Seibold, S. Caprara, M. Grilli, and R. Raimondi, (2017) Journal of Magnetism and
Magnetic Materials, http://dx.doi.org/10.1016/j.jmmm.2016.12.066.

\bibitem{wang04} X. F. Wang, (2004)  Phys. Rev. B {\bf 69}, 035302 .

\bibitem{japa09} G. I. Japaridze, Henrik Johannesson, and Alvaro Ferraz,  (2009) Phys. Rev. B {\bf 80}, 041308(R).

\bibitem{malard11} Mariana Malard, Inna Grusha, G. I. Japaridze, and Henrik Johannesson,  (2011) Phys. Rev. B {\bf 84}, 075466.
                   
\bibitem{rashba-qhe} E. I. Rashba, (1960) Fiz. Tverd. Tela (Leningrad) {\bf 2}, 1224. [E. I. Rashba, (1960) Sov. Phys.-Solid State {\bf 2} 1109]

\bibitem{hofstadter-1976} D. R. Hofstadter,  (1976) Phys. Rev. B {\bf 14}, 2239.

\bibitem{demikhovskii-2006}  V. Ya. Demikhovskii and A. A. Perov, (2006) Europhys. Lett., {\bf 76}, 477

\bibitem{morais-smith-2012} W. Beugeling, N. Goldman, and C. Morais Smith, (2012) Phys. Rev. B {\bf  86}, 075118

\bibitem{venturelli-2011} D. Venturelli {\it et al.}, (2011) Phys. Rev. B {\bf 83}, 075315.

\bibitem{champel-2013} Daniel Hernang\'omez-P\'erez, {\it et al.}, (2013) Phys. Rev. B {\bf 88}, 245433

\bibitem{hashimoto-2008} K. Hashimoto, {\it et al.}, (2008) Phys. Rev. Lett. {\bf 101}, 256802.

\bibitem{lado-2013} J. L. Lado, {\it et al.}, (2013) Phys. Rev. B {\bf 88}, 035448.

\bibitem{giovannetti-2012} Luca Chirolli, Davide Venturelli, Fabio Taddei, Rosario Fazio, and Vittorio Giovannetti, (2012) 
Phys. Rev. B {\bf  85}, 155317


\end{thebibliography}
\end{document}